\documentclass[1p]{elsarticle}

\usepackage{array}
\usepackage{amsmath}
\usepackage{amsbsy}
\usepackage{amssymb}
\usepackage{algorithm}
\usepackage{algpseudocode}
\usepackage{float}
\usepackage{listings}
\usepackage{url}
\usepackage{hyperref}
\usepackage{graphicx}
\usepackage{enumitem}
\usepackage{bm}

\usepackage{tikz-cd}
\usepackage{adjustbox}

\newfloat{algorithm}{tbp}{loa}
\providecommand{\algorithmname}{Algorithm}
\floatname{algorithm}{\protect\algorithmname}

\DeclareRobustCommand{\empirical}[1]{ \hat{#1} }

\newcommand{\Nevents}{M}
\newcommand{\Nsamples}{N}

\newcommand{\op}{\omega}
\newcommand{\opinion}[1]{\omega_{#1}}
\newcommand{\multinomialOp}{\left( \mathbf{b}, u, \mathbf{a} \right)}
\newcommand{\binomialOp}{\left(b, d, u, a \right)}
\newcommand{\multiplication}{\cdot}
\newcommand{\cfusion}{\circledcirc}
\newcommand{\truepdf}[1]{p_{#1}}
\newcommand{\slpdf}[1]{p_{#1}^{SL}}
\newcommand{\empiricalpdf}[1]{\empirical{p}_{#1}}
\newcommand{\empiricalmcpdf}[1]{\empirical{p}_{#1}^\mathrm{MC}}
\newcommand{\empiricalmmpdf}[1]{\empirical{p}_{#1}^\mathrm{MM}}
\newcommand{\empiricalgausspdf}[1]{\empirical{p}_{#1}^\mathrm{GAUSS}}
\newcommand{\empiricalbetapdf}[1]{\empirical{p}_{#1}^\mathrm{BETA}}
\newcommand{\empiricalkdepdf}[1]{\empirical{p}_{#1}^\mathrm{KDE}}
\newcommand{\mmgausspdf}[1]{{p}_{#1}^\mathrm{GAUSS}}
\newcommand{\mmbetapdf}[1]{{p}_{#1}^\mathrm{BETA}}

\newcommand{\SLspace}{\mathcal{S}}
\newcommand{\PDFspace}{\mathcal{P}}
\newcommand{\SLoperator}{\circ_{SL}}
\newcommand{\PDFoperator}{\circ_{P}}

\newcommand{\uniformpdf}[2]{\mathtt{Unif}\left(#1,#2\right)}
\newcommand{\betapdf}[2]{\mathtt{Beta}\left(#1,#2\right)}
\newcommand{\dirichletpdf}[1]{\mathtt{Dir}\left(#1\right)}
\newcommand{\gaussianpdf}[2]{\mathtt{N}\left(#1,#2\right)}

\newcommand{\mom}[2]{M_{#1}\left[#2\right]}
\newcommand{\mean}[1]{E\left[#1\right]}
\newcommand{\var}[1]{Var\left[#1\right]}
\newcommand{\pdfdist}[3]{D_{#1}\left[#2,#3\right]}
\newcommand{\empiricalpdfdist}[3]{\empirical{D}_{#1}\left[#2,#3\right]}
\newcommand{\empiricalmom}[2]{\empirical{M}_{#1}\left[#2\right]}

\newcommand{\TODO}[1]{}

\newcommand{\estimate}{\hat{=}}
\newcommand{\distributes}{\sim}
\newcommand{\domR}{\mathbb{R}}
\newcommand{\domRgrez}{\mathbb{R}_{\ge0}}
\newcommand{\bigO}[1]{\mathcal{O}\left(#1\right)}
\newcommand{\absval}[1]{\left| #1 \right|}

\usepackage[author={FMZennaro}]{pdfcomment}
\newcommand{\rev}[2]{#2}

\makeatletter
\def\ps@pprintTitle{%
	\let\@oddhead\@empty
	\let\@evenhead\@empty
	\def\@oddfoot{}%
	\let\@evenfoot\@oddfoot}
\makeatother

\begin{document}
	
	\begin{frontmatter}
		
		\title{An Empirical Evaluation of the Approximation of Subjective Logic Operators Using Monte Carlo Simulations}
		
		\author[mymainaddress]{Fabio Massimo Zennaro\corref{mycorrespondingauthor}}
		\cortext[mycorrespondingauthor]{Corresponding author}
		\ead{fabiomz@ifi.uio.no}
		
		\author[mymainaddress]{Magdalena Ivanovska}
		\ead{magdalei@ifi.uio.no}
		
		\author[mymainaddress]{Audun J\o sang}
		\ead{josang@mn.uio.no}
		
		\address[mymainaddress]{ Department of Informatics, University of Oslo \\ PO Box 1080 Blindern \\		0316 Oslo, Norway}
		
		\begin{abstract}
			In this paper we analyze the use of subjective logic as a framework for performing approximate transformations over probability distribution functions. As for any approximation, we evaluate subjective logic in terms of computational efficiency and bias. However, while the computational cost may be easily estimated, the bias of subjective logic operators have not yet been investigated. In order to evaluate this bias, we propose an experimental protocol that exploits Monte Carlo simulations and their properties to assess the distance between the result produced by subjective logic operators and the true result of the corresponding transformation over probability distributions. This protocol allows a modeler to get an estimate of the degree of approximation she must be ready to accept as a trade-off for the computational efficiency and the interpretability of the subjective logic framework. Concretely, we apply our method to the relevant case study of the subjective logic operator for binomial multiplication and fusion, and we study empirically their degree of approximation.
		\end{abstract}
		
		
	\end{frontmatter}

\section{Introduction \label{sec:Introduction}}

Subjective logic (SL) \cite{jøsang2016subjective} defines a framework for expressing uncertain probabilistic statements in the form of \textit{subjective opinions}. A subjective opinion allows a modeler to state probabilities over a set of alternative events along with a measure of the global uncertainty of such modeling.
Subjective opinions thus integrate a form of \textit{first-order uncertainty}, relative to the distribution of probability mass over events, and a form of \textit{second-order uncertainty}, due to the incertitude in distributing the probability mass.
Subjective opinions provide a simple, clean and interpretable way to encode and manipulate uncertainty; as such, they constitute a useful modelling tool in sensitive scenarios in which statistical models can not be inferred from data, but must be built relying on the domain knowledge or the intuition of experts. In this fashion, SL has been extensively adopted to model uncertainty in several fields such as trust modeling, biomedical data analysis or forensics analysis \cite{jøsang2016subjective}.

From a purely statistical point of view, subjective opinions can be seen as an alternative representation for standard probability distribution functions (pdfs), such as Beta pdfs or Dirichlet pdfs. Indeed, under certain assumptions, it is possible to define a unique mapping between subjective opinions and probability distribution functions \cite{jøsang2016subjective}. This means that subjective opinions may be interpreted as a re-parametrization of standard distributions from the statistical literature.

SL also defines several operators over subjective opinions. These operators allow to carry out transformations over subjective opinions in a very efficient way. With respect to the underlying probability distributions, SL operators provide an extremely quick approximation of operations over probability distributions that would be otherwise very difficult or impossible to evaluate analytically.

Thus, beyond its original application, SL may also be seen as an effective statistical tool to compute approximate probability distributions generated by the transformations encoded into the SL operators. However, while the efficiency of SL operators may be easily evaluated, estimates about their bias are lacking. This shortcoming may limit the adoption of SL in favor of other better-studied approaches, such as \textit{Monte Carlo (MC) simulations}. Modern \textit{probabilistic programming languages} \cite{ghahramani2015probabilistic} provide a versatile language in which operations over probability distributions may be easily defined and evaluated using pre-coded inference algorithms. While being computationally more expensive, these techniques provide comforting guarantees on the convergence of the algorithms as a function of the number of sampling iterations. These guarantees, contrasting the lack of formal bounds of SL operators, may be a strong argument for many researchers to overlook SL and the related set of operators. 

In this paper, we propose a protocol to address numerically the problem of characterizing the approximation of SL operators by offering an empirical analysis of their bias with respect to \textit{MC simulations}. SL operators and MC simulations are taken as two distinct frameworks to approximate operations over pdfs, each one with its strenghts and limitations. Our analysis defines a quantitative comparison in which SL operators and MC simulations are contrasted in terms of the trade-off between computational efficiency and bias. More specifically, our approach allows to answer the question: \textit{What amount of approximation should we be ready to accept in exchange for the computational efficiency of subjective logic?}

\rev{Edited to take into account constrained fusion}{To show the usefulness} of our protocol, we consider the specific case of \textit{binomial multiplication} and \textit{fusion}. 
Binomial multiplication is a simple SL operator that returns the approximation of the product of two Beta pdfs. Computing the product of independent Beta pdfs is a non-trivial problem \cite{coelho2012distribution} with relevant applications in fields such as reliability analysis and operations research \cite{pham2002product}. Binomial multiplication in SL may then be seen as a simple and effective algorithm to compute an approximate solution to the problem of multiplying together two Beta pdfs. 
Fusion is a SL operator used for merging the opinions of different agents. This operator has been studied and applied in the context of second-order Bayesian networks \cite{kaplan2018efficient}.
For both operators, we compare the approximation obtained using SL to moment-matching approximation and kernel-density approximation produced via MC simulations. In this way, we are able to get an understanding of the amount of approximation that we should be ready to accept if we want to work in the framework of SL.

\TODO{contribs?}

\rev{The outline has been updated with the new Section}{The rest of the paper} is organized as follows. Section \ref{sec:SL} reviews the basics of subjective logic and Section \ref{sec:CompStats} presents the main aspects of computational statistics relevant to this work. Section \ref{sec:Complexity} describes the computational complexity of SL approximations and MC approximations, while Section \ref{sec:Approximation} discusses the bias of the same techniques. Section \ref{sec:Framework} proposes a grounded framework for evaluating the degree of approximation of SL operators in relation to MC simulations. Section \ref{sec:BetaProd} makes this framework concrete by applying it to the case study of the product of Beta pdfs, and it presents a set of empirical simulations to validate our approach; similarly, Section \ref{sec:CFusion} applies our framework to another case study, the fusion of Beta pdfs, and it validates our methodology via empirical simulations. Finally, Section \ref{sec:Conclusion} summarizes the results and discusses possible directions for future work. \rev{New notation table}{For convenience and reference}, Table \ref{tab:Notation} summarizes the notation that will be used throughout this paper.

\begin{table}
	\begin{centering}
\scalebox{0.9}{
\begin{tabular}{cl}
	\hline
	$\Omega$ & Collection of mutually exclusive events \tabularnewline
	$\Nevents$ & Number of mutually exclusive events \tabularnewline
	$X, Y, Z...$ & Random variables over $\Omega$ \tabularnewline
	$x, y, z...$ & Sample of a random variable \tabularnewline
	& \tabularnewline
	
	$\opinion{X}$ & Subjective logic opinion \tabularnewline
	$\mathbf{b}, b$ & Belief (vector and scalar) \tabularnewline
	$d$ & Disbelief (scalar) \tabularnewline
	$\mathbf{a}, a$ & Prior probability (vector, scalar) \tabularnewline
	$u$ & Uncertainty (scalar) \tabularnewline
	& \tabularnewline	
	
	$\truepdf{X}$ & Probability distribution function (pdf) of X \tabularnewline
	$\mom{i}{X}$ & $i$-th moment of the pdf of X \tabularnewline
	$\mean{X}$, $\var{X}$ & Expected value and variance of the pdf of X \tabularnewline
	$\pdfdist{A}{\truepdf{X}}{\truepdf{Y}}$ & Distance $A$ between the pdf of X and the pdf of Y \tabularnewline

	$\empiricalpdf{X}$ & Empirical pdf for X estimated from samples \tabularnewline
	$\Nsamples$ & Number of samples  \tabularnewline
	& \tabularnewline
	
	$\slpdf{X}$ & Pdf underlying a subjective logic opinion\tabularnewline	
	$\empiricalmcpdf{X}$ & Empirical pdf for X estimated via Monte Carlo (MC) sampling \tabularnewline
	$\empiricalkdepdf{X}$ & Empirical pdf for X estimated via MC and kernel density estimation (KDE) \tabularnewline
	$\empiricalmmpdf{X}$ & Empirical pdf for X estimated via MC and moment matching (MM) \tabularnewline
	$\empiricalgausspdf{X}$ & Empirical pdf for X estimated via MC and MM with a Gaussian approximation \tabularnewline
	$\empiricalbetapdf{X}$ & Empirical pdf for X estimated via MC and ad MM with a Beta approximation \tabularnewline
	$\mmgausspdf{X}$ & Pdf for X estimated via analytic MM with a Gaussian approximation \tabularnewline
	$\mmbetapdf{X}$ & Pdf for X estimated via analytic MM with a Beta approximation \tabularnewline
	& \tabularnewline
	
	$\PDFspace$ & Space of probability distribution functions  \tabularnewline
	$\SLspace$ & Space of subjective opinions  \tabularnewline
	$\PDFoperator: \PDFspace \times \PDFspace \rightarrow \PDFspace$ & Binary operator on the space of probability distribution functions  \tabularnewline
	$\SLoperator: \SLspace \times \SLspace \rightarrow \SLspace$ & Binary operator on the space of subjective logic opinions  \tabularnewline
	\hline
\end{tabular}}

\caption{Summary of notation. \label{tab:Notation}}

\end{centering}
\end{table}

\section{Subjective Logic \label{sec:SL}}

In this section, we present the fundamentals of SL. We start with a formalization of subjective opinions and we show how they may be mapped to probability distributions.

\paragraph{Subjective opinions}
Let $\Omega$ be a discrete collection of $\Nevents$ mutually exclusive and exhaustive events. A subjective opinion $\op$ is a triple:
\begin{equation}\label{eq:multinomial_opinion}
\multinomialOp,
\end{equation}
such that
\begin{equation}\label{eq:opinion_constraint}
\sum_{i=1}^{\Nevents} b_i + u = 1,
\end{equation}
where $\mathbf{b} \in \domR^\Nevents$, with $b_i \in \domRgrez$, is the \textit{belief} vector expressing the probability mass that the modeler places on each event $x_i$ in $\Omega$, $u \in \domRgrez$ is the \textit{uncertainty} scalar quantifying the uncertainty of the modeler in its definition of $\mathbf{b}$, and $\mathbf{a} \in \domR^\Nevents$ is the \textit{prior} vector encoding a prior probability distribution over the events in $\Omega$. This subjective opinion is called a \textit{multinomial opinion}.\\
Notice that the constraint in Equation \ref{eq:opinion_constraint} limits the degrees of freedom of $\mathbf{b}$ and $u$ to $\Nevents$ and, consequently, defines a $\Nevents$-dimensional simplex on which subjective opinions may be represented.

The limit-case multinomial opinion is the \textit{binomial opinion} for $\Nevents=2$. In this case $\Omega=\lbrace x, \overline{x} \rbrace $ and the subjective opinion in Equation \ref{eq:multinomial_opinion} may be re-written for simplicity as: 
\begin{equation}
\binomialOp,
\end{equation}
such that
\begin{equation}
b + d + u = 1,
\end{equation}
where $b \in \domRgrez$ is the belief scalar expressing the probability of $x$, $d \in \domRgrez$ is the disbelief scalar expressing the probability of $\overline{x}$, $u \in \domRgrez$ is the uncertainty scalar and $a \in \domRgrez$ is a scalar expressing the prior probability of $x$.\\
Having only two degrees of freedom, binomial opinions in the form $\binomialOp$ belong to a two-dimensional simplex and may be visualized together with $a$ in a barycentric coordinate system\footnote{See \url{http://folk.uio.no/josang/sl/BV.html} for an illustration.}.

\paragraph{Mapping of subjective opinions}

In order to ground SL, a mapping has been defined between multinomial opinions and Dirichlet pdfs and between binomial opinions and Beta pdfs.

\TODO{does it make sense?}
Given a mapping constant $W \in \domRgrez$, it is possible to define a unique mapping from opinions to pdfs. Let $\op = \multinomialOp$ be a multinomial opinion with $u\neq0$; $\op$ can be mapped to a Dirichlet pdf $p$ with distribution $\dirichletpdf{\pmb{\alpha}}$, where the vector of parameters $\pmb{\alpha}$ is defined as:
\begin{equation}
	\pmb{\alpha} = W \left( \frac{\mathbf{b}}{u} + \mathbf{a} \right).
\end{equation}
For binomial opinions, specifically, given a mapping constant $W \in \domRgrez$, it is possible to define a unique mapping from opinions to pdfs. Let $\op = \binomialOp$ be a binomial opinion with $u\neq0$; $\op$ can be mapped to a Beta pdf $p$ with distribution $\betapdf{\alpha}{\beta}$, where $\alpha$ and $\beta$ are parameters defined as:
\begin{equation}
\begin{cases}
\alpha= W \left( \frac{b}{u} + a \right)\\
\beta= W \left( \frac{d}{u} + (1-a) \right).
\end{cases}
\end{equation}
Notice that, for reasons of consistency, $W$ is usually fixed to $2$ \cite{jøsang2016subjective}. We then have a mapping $s$ from opinion $\op$ to pdf $p$:
\[
s: \op \mapsto p.
\]

Vice versa, given a mapping constant $W \in \domRgrez$ and a fixed prior distribution $\mathbf{a}$, it is possible to define a unique mapping from pdfs to opinions. Let $p$ be a Dirichlet pdf with distribution $\dirichletpdf{\pmb{\alpha}}$ with $\alpha_i>1$; $p$ can be mapped to a multinomial opinion $\op = \multinomialOp$, where the parameters are computed as:
\begin{equation}
\begin{cases}
\mathbf{b}= \frac{\pmb{\alpha} - W \mathbf{a}}
{W + \sum_{i} \left( \pmb{\alpha}_i - W \mathbf{a}_i \right)} = 
\frac{\pmb{\alpha} - W \mathbf{a}}{\sum_{i} \pmb{\alpha}_i}\\

\mathbf{u} = \frac{W}
{W + \sum_{i} \left( \pmb{\alpha}_i - W \mathbf{a}_i \right)} = 
\frac{W}{\sum_{i} \pmb{\alpha}_i}\\
\mathbf{a} = \mathbf{a}
\end{cases}
\end{equation}
Again, for a binomial opinion, given a mapping constant $W \in \domRgrez$ and a fixed prior distribution $a$, it is possible to define a unique mapping from pdfs to opinions. Let $p$ be a Beta pdf with distribution $\betapdf{\alpha}{\beta}$ with $\alpha,\beta>1$; $p$ can be mapped to a binomial opinion $\op = \binomialOp$, where the parameters are computed as:
\begin{equation}
\begin{cases}
b= \frac{\alpha - W a}{\alpha + \beta}\\
d= \frac{\beta - W(1-a)}{\alpha + \beta}\\
u = \frac{W}{\alpha + \beta}\\
a = a
\end{cases}
\end{equation}
For reasons of consistency, $W$ is usually fixed to $2$ \cite{jøsang2016subjective}. Given a prior distribution $a$, this generates the mapping $t$ from pdf $p$ to opinion $\op$:
\[
t: p \mapsto \op.
\]

\paragraph{Subjective opinion operators} SL defines several operators over subjective opinions, such as addition, product or fusion \cite{jøsang2016subjective}. In general, these operators are computed over the parameters of subjective opinions. Let $\opinion{X} = \left( \mathbf{b}_X, u_X, \mathbf{a}_X \right)$ and $\opinion{Y} = \left( \mathbf{b}_Y, u_Y, \mathbf{a}_Y \right)$ be two subjective opinions and let $\SLoperator: \SLspace \times \SLspace \rightarrow \SLspace$ be a generic operator over the space of subjective opinions $\SLspace$. Then, $\opinion{Z} = \left( \mathbf{b}_Z, u_Z, \mathbf{a}_Z \right)$ resulting from the application of the operator to $\opinion{X}$ and $\opinion{Y}$ is given as:
\begin{equation}\label{eq:SLoperator}
\opinion{Z} = \opinion{X} \SLoperator \opinion{Y} =
\begin{cases}
\mathbf{b}_Z = f_b \left( \opinion{X}, \opinion{Y} \right)\\

u_Z  = f_u \left( \opinion{X}, \opinion{Y} \right)\\

\mathbf{a}_Z = f_a \left( \opinion{X}, \opinion{Y} \right),
\end{cases}
\end{equation}
where $ f_b, f_u, f_a: \SLspace \times \SLspace \rightarrow \domRgrez $ are operator-specific functions returning the values of belief, uncertainty and prior for the opinion $\opinion{Z}$.

\paragraph{Subjective opinion operators for evaluating operations over pdfs}
When properly defined, SL operators can be used to approximate operations over probability distribution functions. Suppose we are given two pdfs, $p_X$ and $p_Y$, and we want to compute a generic operation over them, $\PDFoperator: \PDFspace \times \PDFspace \rightarrow \PDFspace$ over the space of probability distributions $\PDFspace$. Computing this operation over probability distributions may be very complex. However, if we have an SL operator $\SLoperator: \SLspace \times \SLspace \rightarrow \SLspace$ that approximates $\PDFoperator$, we may find a workaround computing $p_X \PDFoperator p_Y$ by projecting the two distribution onto the opinions $\opinion{X}$ and $\opinion{Y}$, computing the resulting opinion $\opinion{Z} = \opinion{X} \SLoperator \opinion{Y}$, and then mapping the result back onto a probability distribution function $\slpdf{Z}$. In this way, the resulting pdf $\slpdf{Z}$ provides an easy-to-compute approximation of the real pdf $\truepdf{Z}$ (see Figure \ref{fig:SLapproximation}). 

\begin{figure}
	\centering
	\begin{tikzcd}		
		\left(p_X, p_Y\right) \arrow[mapsto]{rr}{\PDFoperator} \arrow[mapsto]{d}{t()} & & \truepdf{Z}  \\
		\left(\opinion{X}, \opinion{Y} \right) \arrow[mapsto]{r}{\SLoperator} &
		\opinion{Z} \arrow[mapsto]{r}{s()}
		& \slpdf{Z} \arrow[dashrightarrow]{u}
	\end{tikzcd}
	\caption{If the application of the operator $\PDFoperator$ to two pdfs $p_X$ and $p_Y$ can not be solved analytically, we can map $p_X$ and $p_Y$ to the opinions $\opinion{X}$ and $\opinion{Y}$ and apply the SL operator $\SLoperator$ to compute the opinion $\opinion{Z}$. The pdf $\slpdf{Z}$ associated with $\opinion{Z}$ provides an approximation of $p_Z$. \label{fig:SLapproximation}}
\end{figure}

\section{Computational Statistics \label{sec:CompStats}}

In this section, we review some elements of computational statistics that are relevant to our work. We describe how sampling is used in MC simulations; we show how unbiased estimators can be built via MC integration; we discuss how unbiased estimators can be used to build moment-matching approximation; we show how pdfs may be reconstructed through kernel density estimation; and, finally, we bring these parts together to show how MC simulations may be used to compute the product of pdfs via moment-matching or kernel-density estimation.

\paragraph{Monte Carlo sampling}
MC simulations are stochastic numerical algorithms designed to find approximate solutions through repeated random sampling. This paradigm has been applied in many areas of research to solve problems whose exact analytical solution is impossible or too difficult to derive. In statistics, MC simulations are widely used to evaluate probability distributions whose analytical form can not be explicitly expressed. 
Let $X$ be a random variable with a probability distribution $p_X$ on the support $\Omega$; let us also assume that the analytical form of $p_X$ is unknown but that we can sample realizations $x_i$ of the random variable $X$; then, MC simulations allow us to draw a large number of independent samples $x_i$ and use them to (i) compute useful empirical statistical descriptors $\hat{S}_X$ of the probability distribution $p_X$, or, eventually, (ii) reconstruct the approximate shape of the probability distribution $p_X$.   

\paragraph{Monte Carlo integration}
In order to compute useful empirical statistical descriptors $S_X$ of the probability distribution $p_X$, MC simulations rely on integration and on the law of large numbers.
Let $S_X$ be a statistics of the probability distribution $p_X$ that can be computed from a function $f\left(\cdot\right)$ applied to the samples $x_i$. The statistics $S_X$ is then defined as:
\begin{equation}
S_X = \int_\Omega f(x) p_X(x) dx.
\end{equation}
By the law of large numbers, an estimator of $S_X$ can be computed using $\Nsamples$ samples of $x_i$ as:
\begin{equation}\label{eq:MCestimator}
\hat{S}_X = \frac{1}{\Nsamples}\sum_{i=1}^\Nsamples f\left(x_i\right).
\end{equation}
It is immediate to see that using Equation \ref{eq:MCestimator} and choosing an appropriate function $f(\cdot)$ we can directly estimate useful statistics of the distribution $p_X$, such as moments and quantiles.
Thus, through a MC simulation we can sample points from $p_X$ and compute informative estimator statistics $\hat{S}_X$.

\paragraph{Moment-matching approximation}
\rev{In this paragraph I added the observation that sometimes moment-matching approximation may be computed analytically.}{A probability distribution} $p_X$ is completely characterized by the collection of all its moments; if we know the parametric form of the function $p_X$ from which we are sampling from, but we ignore the exact value of its parameters, we can compute an estimate $\hat{p}_X$ by setting the moments to the estimated values $\empiricalmom{i}{X}$. In several scenarios of interest it may actually be possible to compute analytically the value of few lower moments $\mom{i}{X}$ of interest (such as, mean and variance); this approach is well-known and it has been used in the study of SL operators as well (see, for instance, \cite{kaplan2013reasoning}).
In general, though, MC simulation and integration provide an empirical and robust way to compute estimators of the $i$-th moments $\empiricalmom{i}{X}$ of a probability distribution $p_X$, even when no exact analytical formula for computing the moments $\mom{i}{X}$ of interest is available.   

\paragraph{Kernel density estimation}
Beyond computing statistics, it is possible to use samples $x_i$ generated in a MC simulation to reconstruct the actual probability distribution $p_X$.
A standard approach to reconstruct a continuous function $p_X$ from a set of finite points $x_i$ is \textit{kernel density estimation} (KDE). Any function may be expressed as a convolution with a kernel function $\kappa(\cdot)$:
\begin{equation}
p_X = \int_\Omega \kappa(x)dx.
\end{equation}
Practically, it is possible to get an empirical approximation using only a finite set of points $x_i$:
\begin{equation}
\hat{p}_X (x) = \frac{1}{\Nsamples w}\sum_{i=1}^\Nsamples \kappa \left( \frac{x-x_i}{w} \right),
\end{equation}
where the kernel $\kappa(\cdot)$ is a symmetric function, like a triangular function or a Gaussian, and $w$ denotes the width of the kernel; empirical rules are available to select an optimal value for this parameter in relation to the number of samples available \cite{principe2010information}. Thus, using the same MC simulation procedure to sample points from $p_X$ it is possible also to estimate an approximate probability distribution $\hat{p}_X$.

\paragraph{Monte Carlo simulation for evaluating operations over pdfs} Suppose we are given two probability distributions, $p_X$ and $p_Y$, and suppose we want to compute the distribution $p_Z$ determined by the application of operation $\PDFoperator: \PDFspace \times \PDFspace \rightarrow \PDFspace$, that is, $p_Z = p_X \PDFoperator p_Y$. If the pdf $p_Z$ can not be computed analytically, MC simulations may be used to sample from $p_Z$ and to estimate a pdf that approximates $\truepdf{Z}$. 
As a first solution, we could rely on the samples $\lbrace z_1, z_2 \dots z_\Nsamples \rbrace$ obtained by sampling from $p_X$ and $p_Y$ to estimate the moments $\empiricalmom{i}{Z}$ and then instantiate a moment-matching approximation $\empiricalmmpdf{Z}$ (see Figure \ref{fig:MMapproximation}). Alternatively, we could use the same samples $\lbrace z_1, z_2 \dots z_\Nsamples \rbrace$ from $p_Z$ to perform a kernel-density estimation and compute the KDE approximation $\empiricalkdepdf{Z}$ (see Figure \ref{fig:KDEapproximation}). Notice that, differently from the SL approximation $\slpdf{Z}$, we decorate the approximations computed via MC simulations $\empiricalmmpdf{Z}$ and $\empiricalkdepdf{Z}$ with a hat to underline that they are empirical statistics.

\begin{figure}
	\centering
	\begin{tikzcd}		
		\left(p_X, p_Y\right) \arrow[mapsto]{rr}{\PDFoperator} \arrow[mapsto]{d}{MCS} & & \truepdf{Z}  \\
		\lbrace z_1, z_2 \dots z_\Nsamples \rbrace \arrow[mapsto]{r}{MCI} &  \empiricalmom{i}{Z} \arrow[mapsto]{r} & \empiricalmmpdf{Z} \arrow[dashrightarrow]{u}
	\end{tikzcd}
	\caption{Moment-matching approximation via MC simulation. If the product of two pdfs $p_X$ and $p_Y$ can not be solved analytically, we can sample, integrate and estimate the pdf $\empiricalmmpdf{Z}$, which provides an approximation of $\truepdf{Z}$. \textit{MCS} stands for MC sampling, \textit{MCI} stands for MC integration. \label{fig:MMapproximation}}
\end{figure}

\begin{figure}
	\centering
	\begin{tikzcd}		
		\left(p_X, p_Y\right) \arrow[mapsto]{r}{\PDFoperator} \arrow[mapsto]{d}{MCS} & \truepdf{Z}  \\
		\lbrace z_1, z_2 \dots z_\Nsamples \rbrace \arrow[mapsto]{r}{KDE} & \empiricalkdepdf{Z} \arrow[dashrightarrow]{u}
	\end{tikzcd}
	\caption{Kernel-density approximation via MC simulation. If the product of two pdfs $p_X$ and $p_Y$ can not be solved analytically, we can sample and estimate the pdf $\empiricalkdepdf{Z}$, which provides an approximation of $\truepdf{Z}$. \textit{MCS} stands for MC sampling, \textit{KDE} stands for kernel-density estimation. \label{fig:KDEapproximation}}
\end{figure}

\section{Computational Complexity \label{sec:Complexity}}

In this section, we discuss and compare the computational complexity of SL operators and MC simulations.
We will evaluate the computational complexity using the $\bigO{\cdot}$ notation as the time complexity of running a given algorithm as a function of its input. 

\paragraph{Subjective logic}
SL operators are defined to be extremely efficient. Indeed, given two opinions $\opinion{X}$ and $\opinion{Y}$ and the generic operator $\SLoperator: \SLspace \times \SLspace \rightarrow \SLspace$, the computation of $\opinion{Z} = \opinion{X} \SLoperator \opinion{Y}$ usually requires only a limited number of function evaluations, as shown in Equation \ref{eq:SLoperator}. The number of evaluations is $2\multiplication \Nevents +1$, where $\Nevents$ is the number of events over which the opinions are defined. Thus, the overall complexity is $\bigO{\Nevents}$: it depends only on the number of events considered, and it is independent of the actual form of the mapped distributions. This makes SL operators an attractive choice especially when working in lower dimensions.  

\paragraph{Monte Carlo simulation}
The MC approach is, by definition, computationally intensive. The computational complexity of a MC simulation scales as a function of the number $\Nsamples$ of samples that must be produced. Each iteration requires random sampling and the execution of all the operations necessary to sample from $\truepdf{Z}$. Overall, the computational complexity of the MC simulation is $\bigO{\Nsamples}$. 
If we are using MC simulations to estimate a moment-matching approximation $\empiricalmmpdf{Z}$, MC integration allows us to compute statistics from the samples generated during the MC simulation with no additional overall computational complexity.
However, if we want to estimate the actual pdf via KDE we have to take into account an increase in the overall computational complexity from the linear order to the quadratic order $\bigO{\Nsamples^2}$. Computing the pdf $\empiricalkdepdf{Z}$ is then a significantly computationally expensive procedure.\\

It is evident that, taking into account computational complexity only, SL operators dominate MC simulations, with or without KDE, especially considering that the number $\Nsamples$ of samples in a MC simulation is required to grow large in order to return reliable results even in low dimensions.

\section{Bias \label{sec:Approximation}}

In this section, we start analyzing the degree of approximation of SL operators and MC simulations.
We will evaluate the degree of approximation in terms of bias of the estimator $\empiricalpdf{Z}$, that is, as the expected value of the difference between the true distribution and the estimated approximation: $\mean{ \truepdf{Z} - \empiricalpdf{Z} }$.

\paragraph{Subjective logic}
The bias of SL operators is dependent on the definition of the specific operator, and a generic theoretical treatment is not possible. Moreover, an analytic study of the bias is not always available for all possible SL operators. In Section \ref{sec:BetaProd} we will consider the case study of the binomial operator for subjective logic and we will analyze more in detail its specific bias.

\paragraph{Monte Carlo simulation}

MC simulations are known to provide asymptotically unbiased estimators. If we estimate a statistics $S_X$ of the pdf $p_X$ using a MC integration as in Equation \ref{eq:MCestimator}, then $\hat{S}_X$ is an asymptotically unbiased estimator, that is, in the limit of infinite samples, it converges to the true quantity it approximates:
\begin{equation}
\lim_{\Nsamples\rightarrow \infty} \hat{S}_X(\Nsamples) = S_X,
\end{equation}
where we made explicit the dependence of $\hat{S}_X$ on the number of samples $\Nsamples$.

If we use MC integration to estimate the moments $\empiricalmom{}{Z}$ for a moment-matching approximation $\empiricalmmpdf{Z}$, the MC simulation provides us with unbiased estimators of the moments; this means that, by increasing the number of samples generated in a MC simulation, we can get arbitrarily close to the true value of the estimated quantity. However, notice that while the estimated moments $\empiricalmom{}{Z}$ are asymptotically unbiased, the $\empiricalmmpdf{Z}$ is biased; this bias is due to the limited set of moments $\empiricalmom{}{Z}$ used to approximate $\truepdf{Z}$.

If we use a MC simulation to estimate the true pdf directly via KDE, the empirical pdf $\empiricalkdepdf{Z}$ is biased. In this case, it is known that the width parameter $w$ of KDE regulates the trade-off between bias and variance. In general, the bias can be shown to be proportional to the width $w$ of the kernel $\kappa(\cdot)$:
\begin{equation}\label{eq:KDEbias}
E_{KDE} \left[ \truepdf{Z} - \empiricalkdepdf{Z}\right]  \propto w^2,
\end{equation}
under the constraint that $w$ can not be reduced to zero, for statistical and computational reasons \cite{principe2010information}. When using a Gaussian kernel, the widely-adopted \emph{Silverman rule} suggests the adoption of a kernel width of the following size:
\begin{equation} \label{eq:Silverman}
w  = 1.06 \hat{\sigma} \frac{1}{\sqrt[5]{\Nsamples}},
\end{equation}
where $\hat{\sigma}$ is the empirical standard deviation computed from the samples:
\begin{equation}
\hat{\sigma} = \sqrt{ \frac{\sum_{i=1}^{\Nsamples} \left( x_i - \hat{\mu} \right)^2 }{\Nsamples-1}},
\end{equation}
where $\hat{\mu}$ is the empirical mean.
It follows, then, that the bias of the KDE approximation is proportional to:
\begin{equation} \label{eq:KDEbias2}
E_{KDE} \left[ \truepdf{Z} - \empiricalkdepdf{Z}\right]  \propto 1.06^2 \hat{\sigma}^2 \Nsamples^{\frac{2}{5}}.
\end{equation}

As said, this bias can never be reduced to zero. However, in specific computational setting, this bias may be bounded by finding an optimal trade-off between the number of samples $\Nsamples$ and the empirical standard deviation $\hat{\sigma}$. 
In particular, if the domain of $\truepdf{Z}$ is a discrete domain, as in the case of multinomial opinions and Dirichlet pdfs which underlie subjective opinions, then the empirical standard deviation $\hat{\sigma}$ may be bounded and it may be possible to estimate the magnitude of the bias as a function of the number of samples $\Nsamples$.\\ 

In summary, from the point of view of approximation, MC simulations represents a safer choice than SL operators, as they are grounded in solid theory and they allow us to quantify and to control the bias. The lack of any bound for SL operators may be seen as an obstacle in adopting them when working in critical domains where precise approximations are required. In the next section, we will introduce our protocol to solve this problem and estimate the degree of approximation of SL operators.

\section{Computational Evaluation of the Degree of Approximation of Subjective Logic Operators \label{sec:Framework}}

In this section we present a framework to evaluate the bias of an SL operator. We start by discussing how MC approximations may be related to SL approximation using a distance measure; then, we define what precise distance measure we will use and how it relates to bias.

\begin{figure}
	\centering
	\begin{tikzcd}
		\left(\opinion{X}, \opinion{Y} \right) \arrow[mapsto]{r}{\SLoperator} 
		& \opinion{Z} \arrow[mapsto]{r}{s()} & \slpdf{Z}  \arrow[dashrightarrow]{d} \\		
		\left(p_X, p_Y\right) \arrow[mapsto]{rr}{\PDFoperator} \arrow[mapsto]{u}{t()} \arrow[mapsto]{d}{MCS} & & \truepdf{Z}  \\
		\lbrace z_1, z_2 \dots z_\Nsamples \rbrace \arrow[mapsto]{d}{MCI} \arrow[mapsto]{rr}{KDE} &  &
		\empiricalkdepdf{Z} \arrow[dashrightarrow]{u} \\
		\empiricalmom{i}{Z} \arrow[mapsto]{rr} & & \empiricalmmpdf{Z} \arrow[dashrightarrow, bend right]{uu} \\
	\end{tikzcd}
	\caption{Approximations of $\truepdf{Z}$. \textit{MCS} stands for MC sampling, \textit{MCI} stands for MC integration, \textit{KDE} stands for kernel-density estimation. \label{fig:Allapproximation}}
\end{figure}
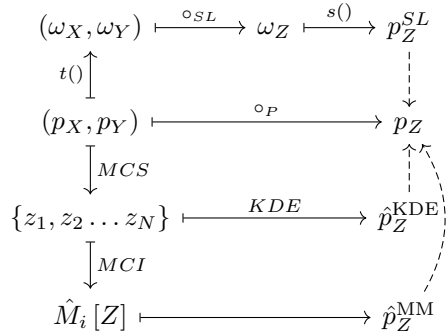

\paragraph{Relating subjective logic approximation and Monte Carlo approximations via a distance measure}
In the previous sections we illustrated two methodologies for finding an approximation of the pdf $\truepdf{Z}$, one based on SL operators ($\slpdf{Z}$) and one relying on MC simulations ($\empiricalkdepdf{Z}, \empiricalmmpdf{Z}$). Figure \ref{fig:Allapproximation} merges the graphs in Figure \ref{fig:SLapproximation}, \ref{fig:MMapproximation} and \ref{fig:KDEapproximation} to illustrate the alternative computational paths that are offered to compute an approximation of $\truepdf{Z}$; starting from the distributions $\left(p_X, p_Y\right)$, the upper path represents the SL approach to finding an approximation of $\truepdf{Z} = p_X \PDFoperator p_Y$, while the lower paths represent MC approaches to finding an approximation of the same quantity $\truepdf{Z}$.

Now, approximate methods trade off precision in the results for simplicity in computation. In order to make a grounded decision on which approximation path in Figure \ref{fig:Allapproximation} to use, it is necessary to quantify the trade-off between computational complexity and bias. As discussed in Section \ref{sec:Complexity} and \ref{sec:Approximation}, in the case of KDE approximation via MC simulations, both complexity and bias are known. However, in the case of SL operators, we may easily derive their computational complexity, but we have no simple way of evaluating their bias.
Exploiting the properties of MC integration and the idea of distance between pdfs, it is possible to assess the degree of approximation of SL operators in a computational fashion by relating them to MC simulations.

A simple way to evaluate how well a pdf $p$ approximates another pdf $q$ is to estimate the distance between them, $\pdfdist{}{p}{q}$, where $\pdfdist{}{\cdot}{\cdot}$ is a measure of distance or divergence between pdfs \cite{mackay2003information}. 
The degree of approximation of $\slpdf{Z}$ could then be obtained by measuring the distance from the true pdf $\truepdf{Z}$:
\begin{equation}\label{eq:dist_true_sl}
\pdfdist{}{\truepdf{Z}}{\slpdf{Z}}.
\end{equation}
However, since the true pdf $\truepdf{Z}$ is taken to be unknown or hard to compute, it is challenging to get a direct estimate of these quantities. Since we can not rely directly on $\truepdf{Z}$, we can instead exploit MC simulations and its properties.

From Equation \ref{eq:KDEbias} in Section \ref{sec:Approximation}, we know that the KDE estimation $\empiricalkdepdf{Z}$ is biased and we know how to evaluate it. Moreover, from Equation \ref{eq:KDEbias2} in Section \ref{sec:Approximation}, we see that this bias depends on the number of samples $\Nsamples$ and the standard deviation $\hat{\sigma}$.
Now, if the domain of $\truepdf{Z}$ is a discrete domain, as in the case of multinomial opinions and Dirichlet pdfs, then the empirical standard deviation $\hat{\sigma}$ may be bounded and it may be possible to estimate the magnitude of the bias as a function of the number of samples $\Nsamples$. It may be possible to select a number of samples $\Nsamples$ that shrinks the bias to a negligible quantity; in such case, we can then accept the KDE estimation $\empiricalkdepdf{Z}$ as a close approximation of the true pdf $\truepdf{Z}$:
\begin{equation}\label{eq:approximationTrueMC}
\pdfdist{}{\truepdf{Z}}{\empiricalkdepdf{Z}(\Nsamples)} \approx 0.
\end{equation}
We underline that this approximation holds only under the assumption that, for an increasing number of samples $\Nsamples$, the bias of the estimate $\empiricalkdepdf{Z}(\Nsamples)$ tends, if not to zero, to a quantity whose order of magnitude is negligible with respect to further analysis; in other words, the validity of the approximation in Equation \ref{eq:approximationTrueMC} is conditional on the pdf $\truepdf{Z}$ we are considering, the analysis we will be carrying out, and the number of samples we can produce (for an example of an evaluation of these conditions, see the application to the case study of the product of Beta pdfs in Section \ref{sec:BetaProd} and Section \ref{sec:EmpircalStudy}).

The approximation in Equation \ref{eq:approximationTrueMC} is extremely useful because it means that while we can not evaluate absolute distances with respect to the true distribution $\truepdf{Z}$, we can still evaluate the relative distance between the KDE approximation and the SL approximation, and use it as a proxy for the distance between the SL approximation $\slpdf{Z}$ and the true distribution $\truepdf{Z}$:
\begin{equation}
\pdfdist{}{\empiricalkdepdf{Z}(\Nsamples)}{\slpdf{Z}} \approx \pdfdist{}{\truepdf{Z}}{\slpdf{Z}}.
\end{equation}
Thus, given only a finite set of samples $\Nsamples$ we can obtain an empirical statistic of the distance as:
\begin{equation}
\empiricalpdfdist{}{\truepdf{Z}}{\slpdf{Z}} \estimate \pdfdist{}{\empiricalkdepdf{Z}(\Nsamples)}{\slpdf{Z}},
\end{equation}
If the condition in Equation \ref{eq:approximationTrueMC} holds, we expect the distance $\pdfdist{}{\empiricalkdepdf{Z}(\Nsamples)}{\slpdf{Z}}$ to be orders of magnitudes greater than $\pdfdist{}{\truepdf{Z}}{\empiricalkdepdf{Z}(\Nsamples)}$; this would indeed confirm that the bias of $\empiricalkdepdf{Z}(\Nsamples)$ is negligible and that the computation of $\empiricalpdfdist{}{\truepdf{Z}}{\slpdf{Z}}$ (using a finite number of samples) provides a good estimate of the degree of approximation of the SL approximation.

\paragraph{Relating distance measure to bias} So far, we have discussed distance measures in abstract terms. The quantity $\empiricalpdfdist{}{\truepdf{Z}}{\slpdf{Z}}$ may indeed be computed using different pdf distance, such as $\phi$-divergences or integral probability metrics \cite{sriperumbudur2009integral}. 

In this paper, we will rely on computing a simple integral distance, defined as:
\begin{equation}
\pdfdist{I}{p}{q} = \int_{-\infty}^{+\infty} \absval{p(x)-q(x)} dx.
\end{equation}

This distance $\pdfdist{I}{p}{q}$ is the same as the \textit{total variation distance} except for the scaling constant:
\begin{equation}
\pdfdist{TV}{p}{q} = \frac{1}{2} \int_{-\infty}^{+\infty} \absval{p(x)-q(x)} dx.
\end{equation}  
The constant $\frac{1}{2}$ rescales the distance on the interval $[0,1]$. However, in order to get an absolute evaluation of how the mass of the two distributions $p$ and $q$ overlaps, we drop the scaling constant.

The choice of an integral distance $\pdfdist{I}{\cdot}{\cdot}$ is justified for three reasons. 
First, from a conceptual point of view, an integral distance allows us to get a complete picture of the difference between two pdfs. While measures based on the evaluation of a limited set of synthetic statistics such as moments would provide us with a rough evaluation of the difference between two distributions, an integral distance provides a more precise way to assess the distribution of the mass of probability, taking into account, for instance, the potential presence of multiple modes or how mass subtly distributes on the tails.

Second, from a computational point of view, the integral distance $\pdfdist{I}{\cdot}{\cdot}$ allows us, once again, to exploit MC integration. Recall that we want to get an estimation of $\empiricalpdfdist{}{\truepdf{Z}}{\slpdf{Z}}$ through the approximation $\pdfdist{}{\empiricalkdepdf{Z}(\Nsamples)}{\slpdf{Z}}$. Now, if we reconstruct $\empiricalkdepdf{Z}$ via KDE, we can estimate the integral distance via MC integration over the domain $\Omega$ of the events as:
\begin{equation}\label{eq:TVDviaMC}
\int_\Omega \absval{ \empiricalkdepdf{Z}(z) - \slpdf{Z}(z) } dz \estimate \frac{1}{\Nsamples} \sum_{i=1}^{\Nsamples} \absval{ \empiricalkdepdf{Z}(z_i) - \slpdf{Z}(z_i) }.
\end{equation}

Third, from a theoretical point of view, the integral in Equation \ref{eq:TVDviaMC} is related to the bias:
\begin{align}
\int_\Omega \absval{ \empiricalkdepdf{Z}(z) - \slpdf{Z}(z) } dz & \estimate \frac{1}{\Nsamples} \sum_{i=1}^{\Nsamples} \absval{ \empiricalkdepdf{Z}(z_i) - \slpdf{Z}(z_i) }\\
& \estimate \mean{\absval{ \empiricalkdepdf{Z}(Z) - \slpdf{Z}(Z)}}\\
& \geq \mean{ \empiricalkdepdf{Z}(Z) - \slpdf{Z}(Z)}.
\end{align}
Thus, using the integral distance $\pdfdist{I}{\empiricalkdepdf{Z}}{\slpdf{Z}}$ we can obtain an estimation of the distance $\empiricalpdfdist{}{\truepdf{Z}}{\slpdf{Z}}$ as well as an upper bound on the bias of $\slpdf{Z}$. Notice that the absolute value in the integral distance provides a more honest evaluation of the absolute difference between pdfs, avoiding an averaging effect in absence of the absolute value operator.

\begin{figure}
	\centering
	\begin{tikzcd}
		\left(\opinion{X}, \opinion{Y} \right) \arrow[mapsto]{r}{\SLoperator} 
		& \opinion{Z} \arrow[mapsto]{r}{s()} & \slpdf{Z} \arrow[mapsto]{r} & \int \absval{\slpdf{Z} - \empiricalkdepdf{Z}} \arrow[mapsto]{r}{MCI} & \empiricalpdfdist{I}{\slpdf{Z}}{\truepdf{Z}}\\		
		\left(p_X, p_Y\right) \arrow[mapsto]{rr}{\PDFoperator} \arrow[mapsto]{u}{t()} \arrow[mapsto]{d}{MCS} & & \truepdf{Z} \arrow[dashrightarrow]{urr} \arrow[dashrightarrow]{ddrr} \\
		\lbrace z_1, z_2 \dots z_\Nsamples \rbrace \arrow[mapsto]{d}{MCI} \arrow[mapsto]{rr}{KDE} &  &
		\empiricalkdepdf{Z} \arrow[mapsto]{uur} \arrow[mapsto]{dr} \\
		\empiricalmom{i}{Z} \arrow[mapsto]{rr} & & \empiricalmmpdf{Z} \arrow[mapsto]{r} & \int \absval{\empiricalmmpdf{Z} - \empiricalkdepdf{Z}} \arrow[mapsto]{r}{MCI} & \empiricalpdfdist{I}{\empiricalmmpdf{Z}}{\truepdf{Z}} \\
	\end{tikzcd}
	\caption{Evaluations of the distances between $\truepdf{Z}$ and its approximations based on the assumption that $\pdfdist{}{\truepdf{Z}}{\empiricalkdepdf{Z}(\Nsamples)} \approx 0$. \textit{MCS} stands for MC sampling, \textit{MCI} stands for MC integration, \textit{KDE} stands for kernel-density estimation. \label{fig:ApproximationWithDistances}}
\end{figure}
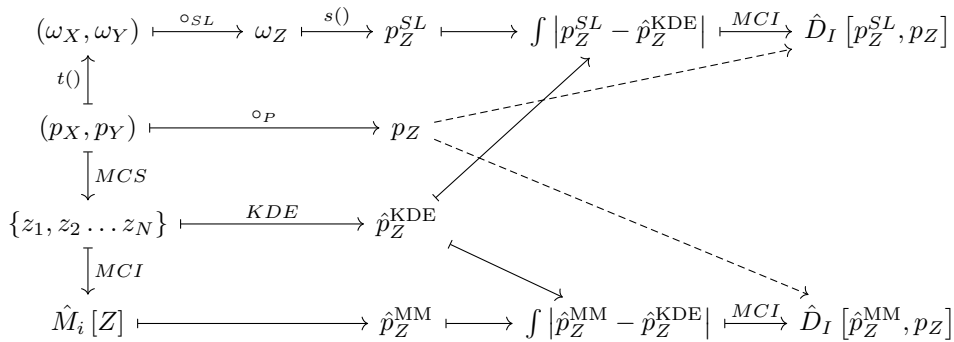

Figure \ref{fig:ApproximationWithDistances} summarizes our overall framework to evaluate the degree of approximation of the SL approximation as the integral distance $\int \absval{\slpdf{Z} - \empiricalkdepdf{Z}}$, under the assumption that $\pdfdist{}{\truepdf{Z}}{\empiricalkdepdf{Z}(\Nsamples)} \approx 0$.
This approach is generic and it is not tied to the SL approximation. If the condition in Equation $\ref{eq:approximationTrueMC}$ can be guaranteed, the same approach may be used to get an estimation of the distance between the true pdf $\truepdf{Z}$ and other potential approximation. For instance, Figure \ref{fig:ApproximationWithDistances} shows our methodology applied also to the problem of estimating the distance from the true pdf of the moment-matching approximation $\pdfdist{}{\truepdf{Z}}{\empiricalmmpdf{Z}(\Nsamples)}$ by computing the distance $\int \absval{\empiricalmmpdf{Z} - \empiricalkdepdf{Z}}$.

\section{Case Study: Product of Beta Distributions \label{sec:BetaProd}}

In this section, we show how our framework may be applied to the problem of computing the product of Beta distributions. We first recall the definition of a Beta distribution and the definition of the product of Beta distributions; we then introduce the SL operator for binomial multiplication and we discuss how it can be used for approximating the distribution of the random variable given by the product of two independent random variables with Beta distributions; we work out the computational complexity of binomial multiplication and show the lack of generic estimate of its degree of approximation; to solve this problem, we apply our framework to get an evaluation of the degree of approximation of binomial multiplication; finally, we run an extensive set of empirical simulations to validate our theoretical results. 

\paragraph{Beta pdf} Let $X$ be a random variable on the support $[0,1]$; we say that $X$ follows a Beta distribution $X\distributes \betapdf{\alpha}{\beta}$ with parameters $\alpha \in \domRgrez$ and $\beta \in \domRgrez$ when its probability density function $p_X$ has the following form:
\begin{equation}
p_X\left(x;\alpha,\beta\right) =
\frac{1}{B\left(\alpha,\beta\right)}
x^{\alpha-1} \left(1-x\right)^{\beta-1},
\end{equation}
where $B\left(\alpha,\beta\right)$ is the Beta function.

\paragraph{Product of Beta pdfs} Let $X \distributes \betapdf{\alpha_X}{\beta_X}$ and $Y \distributes \betapdf{\alpha_Y}{\beta_Y}$ be two independent Beta random variables with associated pdfs $p_X$ and $p_Y$. Let us define a third random variable $Z$ as the product of the two Beta random variables $Z = X \multiplication Y$. The probability density function $p_Z$ of $Z$ does not follow a Beta distribution anymore, and its precise analytical form can not be easily expressed using elementary functions \cite{pham1994reliability}.
An analytical solution to the evaluation of the pdf of the product of two Beta distributions has been offered in \cite{pham2002product}\footnote{This paper actually presents  the more generic solution to the problem of multiplying two \textit{general Beta distribution}, which subsume the multiplication of two simple Beta distributions as defined above.}:
\begin{equation}
\begin{array}{c}
p_{Z}\left(z;\alpha_{X},\alpha_{Y},\beta_{X},\beta_{Y}\right)=\\
B\left(\beta_{X},\beta_{Y}\right)\cdot z^{-\beta_{X}}\cdot\left(1-z\right)^{\beta_{X}+\beta_{Y}-1}\cdot z^{\alpha_{Y}}\cdot\\
\cdot\frac{F_{D}^{(3)}\left(\beta_{X};1-\alpha_{X},1-\alpha_{Y},\alpha_{X}+\beta_{X}-1;\beta_{X}+\beta_{Y};0,\frac{z-1}{z},\frac{z-1}{z}\right)}{B\left(\alpha_{X},\beta_{X}\right)B\left(\alpha_{Y},\beta_{Y}\right)},
\end{array}
\end{equation}
where $F_{D}^{(3)}$ is the Lauricella D hyper-geometric series. \TODO{Insert definition?}
While this formula provides an elegant solution to the problem of finding the pdf of the product of two Beta pdfs, its straightforward evaluation is challenging as the Lauricella function requires the computation of factorial products and series.

Other analytical approaches to evaluate the product of two or more Beta distributions have been proposed, including methods relying on high-order functions, such as the Meijer G-function or Fox's H function, or modeling the pdf of the product using an infinite mixture of simpler distributions \cite{tang1984distribution, coelho2012distribution}. These approaches also present computational challenges, despite more efficient solutions have been investigated \cite{tang1984distribution, coelho2012distribution}. 

Finally, common approaches rely on MC simulations to sample points from the probability distribution of $Z$ and to compute statistics of the pdf by matching the moments or the quantiles of $Z$ \cite{pham1994reliability}, as we reviewed in Section \ref{sec:CompStats}. \rev{I added this part on the possibility of using an analytic approach}{Notice that} when considering the product of two independent random variables $Z = X \multiplication Y$, it is straightforward to compute the mean $\mean{Z}$ and the variance $\var{Z}$ of $\truepdf{Z}$ analytically as:
\begin{equation}\label{eq:analyticMM}
\begin{array}{c}
\mean{Z} = \mean{X} \multiplication \mean{Y}\\
\var{Z} = \mean{X}^2 \multiplication \var{Y} + \mean{Y}^2 \multiplication \var{X} + \var{X}\multiplication\var{Y}.
\end{array}
\end{equation}
Thus, if we were to perform a moment-matching approximation considering only the first two moments, we could instantiate such an approximation without running any MC simulation with a constant computational complexity of $\bigO{1}$.

\paragraph{Subjective logic binomial multiplication}
An alternative solution to compute the product of Beta distributions is based on the use of the SL operator for binomial multiplication.
Given two binomial opinions $\opinion{X}$ and $\opinion{Y}$ defined on different domains, the binomial opinion $\opinion{Z}$ resulting from the multiplication $\opinion{X} \multiplication \opinion{Y}$ is computed as \cite{jøsang2016subjective}:
\begin{equation}
\opinion{Z} = 
\begin{cases}
b_Z = b_X b_Y + \frac{ \left(1-a_X\right) a_Y b_X u_Y + a_X \left(1-a_Y\right) u_X b_Y } {1 - a_X a_Y}\\

d_Z = d_X + d_Y - d_X d_Y\\

u_Z  = u_X u_Y + \frac{ \left(1-a_Y\right) b_X u_Y + (1-a_X) u_X b_Y } {1 - a_X a_Y}\\

a_Z = a_X a_Y.
\end{cases}
\end{equation}
Practically, a binomial product operator allows us to evaluate the combination of two opinions over two different facts.
In the domain of probability distributions, the multiplication of opinions $\opinion{Z} = \opinion{X} \multiplication \opinion{Y}$ translates into the multiplication of the mapped pdfs $p_Z = p_X \multiplication p_Y$.

\paragraph{Approximating the product of Beta pdfs}
Now, assume we are interested in computing the product $Z = X \multiplication Y$, where $X$ and $Y$ are two independent Beta random variables. Since an analytic solution is hard to compute, we may decide to rely either on the SL approximation or on a MC approximation.

Concerning moment-matching approximations we may consider a Gaussian pdf and a Beta pdf. Using a Gaussian pdf is a choice motivated by the simplicity and the ubiquity of this distribution; however, this is clearly a naive choice, as a Gaussian pdf has an unbounded support, is symmetrical and it assumes that all the statistical moments greater than the second are zero. Using a Beta distribution is a more prudent choice: even if it is known that the product of two Betas is not, in general, a Beta distribution, a Beta pdf still fits the right support and it may have other moments different from zero.
In order to evaluate the parameters of our Gaussian and Beta approximation, we may rely on MC simulations or on an analytic evaluation. If we opt for MC simulations, we can use MC integration to estimate the mean $\hat{\mu}$ and the variance $\hat{\sigma}^2$ of $\truepdf{Z}$; the Gaussian approximation $\empiricalgausspdf{Z}$ is then instantiated as $\gaussianpdf{\hat{\mu}}{\hat{\sigma}^2}$, while the Beta approximation $\empiricalbetapdf{Z}$ is defined as $\betapdf
{\frac
	{ -\hat{\mu} \left(\hat{\sigma}^2 + \hat{\mu}^2 - \hat{\mu} \right)  } {\hat{\sigma}^2}}   
{\frac
	{ \left(\hat{\mu}-1\right) \left(\hat{\sigma}^2 + \hat{\mu}^2 - \hat{\mu} \right)  } 
	{\hat{\sigma}^2}}$, thus guaranteeing that $\truepdf{Z}$, $\empiricalgausspdf{Z}$ and $\empiricalbetapdf{Z}$ have the same mean and variance.
\rev{I added here the discussion about the possibility of computing moment-matching approximation analytically}{If we rely on an analytic approach,} we can easily compute the mean $\mu$ and the variance $\sigma^2$ of $\truepdf{Z}$ using Equation \ref{eq:analyticMM}; as before, the Gaussian approximation $\mmgausspdf{Z}$ is then instantiated as $\gaussianpdf{{\mu}}{{\sigma}^2}$, while the Beta approximation $\mmbetapdf{Z}$ is defined as $\betapdf
{\frac
	{ -{\mu} \left({\sigma}^2 + {\mu}^2 - {\mu} \right)  } {{\sigma}^2}}   
{\frac
	{ \left({\mu}-1\right) \left({\sigma}^2 + {\mu}^2 - {\mu} \right)  } 
	{{\sigma}^2}}.$

\begin{figure}
	\centering
	\begin{tikzcd}
		\left(\opinion{X}, \opinion{Y} \right) \arrow[mapsto]{r}{\multiplication} 
		& \opinion{Z} \arrow[mapsto]{r}{s()} & \slpdf{Z}  \arrow[dashrightarrow]{d} \\		
		\left(p_X, p_Y\right) \arrow[mapsto]{rr}{\multiplication}
		\arrow[mapsto, bend right=60, "AMC",swap]{dddd} \arrow[mapsto]{u}{t()} \arrow[mapsto]{d}{MCS} & & \truepdf{Z}  \\
		\lbrace z_1, z_2 \dots z_\Nsamples \rbrace  \arrow[mapsto]{d}{MCI} \arrow[mapsto]{rr}{KDE} &  &
		\empiricalkdepdf{Z} \arrow[dashrightarrow]{u} \\
		\left( \hat{\mu}, \hat{\sigma} \right) \arrow[mapsto]{rr} \arrow[mapsto]{drr} & & \empiricalgausspdf{Z} \arrow[dashrightarrow, bend right]{uu} \\
		& & \empiricalbetapdf{Z} \arrow[dashrightarrow, bend right]{uuu} \\
		\left( {\mu}, {\sigma} \right) \arrow[mapsto]{rr} \arrow[mapsto]{drr} & & \mmgausspdf{Z} \arrow[dashrightarrow, bend right]{uuuu}\\
		& & \mmbetapdf{Z} \arrow[dashrightarrow, bend right]{uuuuu}
	\end{tikzcd}
	\caption{Approximations of the product of two Beta pdfs, $p_X$ and $p_Y$. \textit{MCS} stands for MC sampling, \textit{MCI} stands for MC integration, \textit{KDE} stands for kernel-density estimation, \textit{AMC} stands for analytic moment computation. \label{fig:ApproximationMultiplication}}
\end{figure}
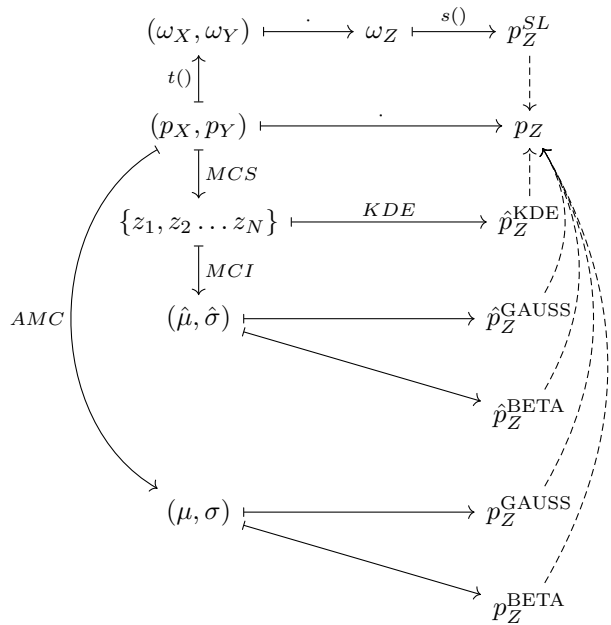

Figure \ref{fig:ApproximationMultiplication} provides a concrete instantiation of the diagram in Figure \ref{fig:Allapproximation}, in which the generic operators $\PDFoperator$ and $\SLoperator$ have been substituted with multiplication and the generic moment-matching approximation $\empiricalmmpdf{Z}$ has been replaced by the empirical Gaussian approximation $\empiricalgausspdf{Z}$, the empirical Beta approximation $\empiricalbetapdf{Z}$, the analytic Gaussian approximation $\mmgausspdf{Z}$, and the analytic Beta approximation $\mmbetapdf{Z}$.

As discussed earlier, choosing which path to take, whether to follow the SL approximation path in upper part of the graph or opt for one of the MC approximations in the lower part, requires evaluating the trade-off between computational complexity and degree of approximation of the different approaches. As these parameters are known in the case of MC simulations, we will review here the computational complexity and the approximation of the binomial multiplication.

\paragraph{Computational complexity of the binomial product operator}
Binomial multiplication is extremely efficient. Given two binomial opinions $\opinion{X}$ and $\opinion{Y}$ it is possible to compute their product $\opinion{Z}$ through a fixed and finite number of arithmetic operations. Independently from the actual form of the mapped distributions, the product is always computed in the same amount of time. As such, the computational complexity of these SL operators is constant $\bigO{1}$. 

\paragraph{Approximation of the binomial operator}
The original paper that introduced the SL operator for binomial multiplication \cite{josang2005multiplication} proposed a first qualitative analysis of the degree of approximation of this operator. In particular, it considered the specific instance of the multiplication of two Beta pdfs of the form $X,Y \distributes \betapdf{1}{1}$; the pdf of $X$ and $Y$ reduces to a uniform distribution over $[0,1]$, which is taken to be a worst-case scenario with maximal variance and entropy. The analytical solution $Z=X \multiplication Y$ to this particular case was then computed and graphically compared to the pdf associated with product $\opinion{Z} = \opinion{X} \multiplication \opinion{Y}$. This study provided a clear visual appraisal of the difference between the exact pdf and the SL-approximated pdf, but no quantitative estimation were provided for more general cases.

\paragraph{Relating the binomial multiplication and Monte Carlo approximations}
In order to compute a numerical estimation of the degree of approximation of the SL operator for binomial multiplication we want to rely on the framework described in Section \ref{sec:Framework}.

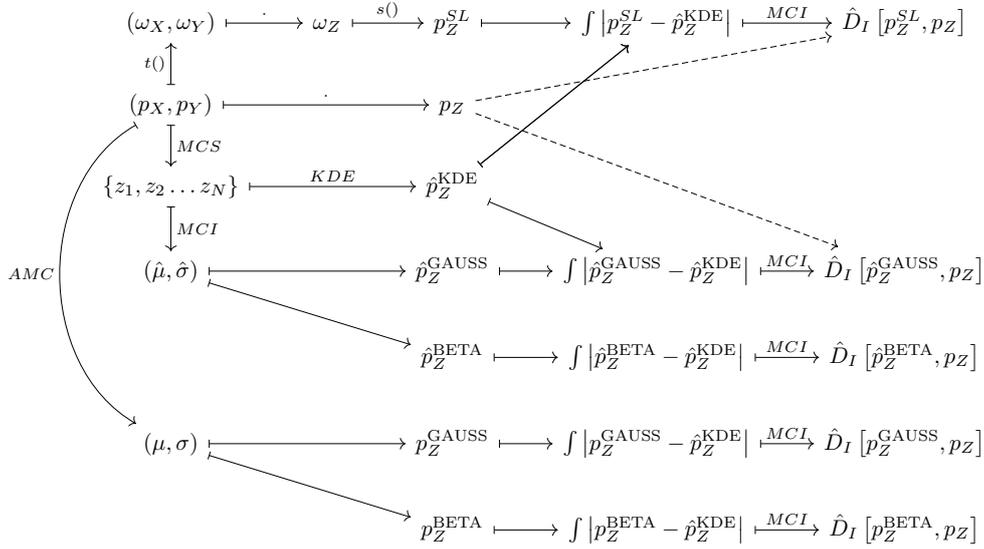
\begin{figure}
	\centering
	\adjustbox{scale=.85}{
	\begin{tikzcd}
		\left(\opinion{X}, \opinion{Y} \right) \arrow[mapsto]{r}{\multiplication} 
		& \opinion{Z} \arrow[mapsto]{r}{s()} & \slpdf{Z} \arrow[mapsto]{r} & \int \absval{\slpdf{Z} - \empiricalkdepdf{Z}} \arrow[mapsto]{r}{MCI} & \empiricalpdfdist{I}{\slpdf{Z}}{\truepdf{Z}}\\		
		\left(p_X, p_Y\right) \arrow[mapsto]{rr}{\multiplication}
		\arrow[mapsto, bend right=60, "AMC",swap]{dddd} \arrow[mapsto]{u}{t()} \arrow[mapsto]{d}{MCS} & & \truepdf{Z} \arrow[dashrightarrow]{urr} \arrow[dashrightarrow]{ddrr} \\
		\lbrace z_1, z_2 \dots z_\Nsamples \rbrace  \arrow[mapsto]{d}{MCI} \arrow[mapsto]{rr}{KDE} &  &
		\empiricalkdepdf{Z} \arrow[mapsto]{uur} \arrow[mapsto]{dr} \arrow[mapsto]{uur}\\
		\left( \hat{\mu}, \hat{\sigma} \right) \arrow[mapsto]{rr} \arrow[mapsto]{drr} & & \empiricalgausspdf{Z} \arrow[mapsto]{r} & \int \absval{\empiricalgausspdf{Z} - \empiricalkdepdf{Z}} \arrow[mapsto]{r}{MCI} & \empiricalpdfdist{I}{\empiricalgausspdf{Z}}{\truepdf{Z}} \\
		& & \empiricalbetapdf{Z} \arrow[mapsto]{r} & \int \absval{\empiricalbetapdf{Z} - \empiricalkdepdf{Z}} \arrow[mapsto]{r}{MCI} & \empiricalpdfdist{I}{\empiricalbetapdf{Z}}{\truepdf{Z}}\\
		\left( {\mu}, {\sigma} \right) \arrow[mapsto]{rr} \arrow[mapsto]{drr} & & \mmgausspdf{Z} \arrow[mapsto]{r} & \int \absval{\mmgausspdf{Z} - \empiricalkdepdf{Z}} \arrow[mapsto]{r}{MCI} & \empiricalpdfdist{I}{\mmgausspdf{Z}}{\truepdf{Z}} \\
		& & \mmbetapdf{Z} \arrow[mapsto]{r} & \int \absval{\mmbetapdf{Z} - \empiricalkdepdf{Z}} \arrow[mapsto]{r}{MCI} & \empiricalpdfdist{I}{\mmbetapdf{Z}}{\truepdf{Z}}\\
	\end{tikzcd}}
	\caption{Evaluations of the distances between $\truepdf{Z}$ and its approximations based on the assumption that $\pdfdist{}{\truepdf{Z}}{\empiricalkdepdf{Z}(\Nsamples)} \approx 0$. \textit{MCS} stands for MC sampling, \textit{MCI} stands for MC integration, \textit{KDE} stands for kernel-density estimation, \textit{AMC} stands for analytic moment computation. \label{fig:ApproximationMultiplicationWithDistances}}
\end{figure}

The basic condition expressed in Equation \ref{eq:approximationTrueMC} requires the bias of $\empiricalkdepdf{Z}$ to be bounded and negligible.
Recall that this bias, using a Gaussian kernel with width computed using the Silverman rule, is $E_{KDE} \left[ \truepdf{Z} - \empiricalkdepdf{Z}\right]  \propto 1.06^2 \hat{\sigma}^2 \Nsamples^{\frac{2}{5}}$, where
\begin{equation}
\hat{\sigma} = \sqrt{ \frac{\sum_{i=1}^{\Nsamples} \left( x_i - \hat{\mu} \right)^2 }{\Nsamples-1}}.
\end{equation}
Now, notice that on our bounded support $\left[0,1\right]$ we can expect the difference $\left( x_i - \hat{\mu} \right)$ to, be at most, in the order of $10^{-1}$. This implies that, in the worst case, the order of magnitude of $\hat{\sigma}$ may be estimated as:
\begin{equation}
\hat{\sigma} < \sqrt{ \frac{ {\Nsamples} \left( 10^{-1} \right)^2 }{\Nsamples-1}}
\end{equation}
\begin{equation}
\hat{\sigma} < 10^{-1}.
\end{equation}
Consequently, relying on Silverman rule in Equation \ref{eq:Silverman}, the order of magnitude of largest kernel width $w$ may be bounded as:
\begin{equation}
w < 1.06 \multiplication 10^{-1} \frac{1}{\sqrt[5]{\Nsamples}}
\end{equation}
\begin{equation}
w < 10^{-1} \Nsamples^{-\frac{1}{5}}.
\end{equation}
As such, from Equation \ref{eq:KDEbias2}, the bias will be proportional to this upper bound:
\begin{equation}\label{eq:kde_bias}
E_{KDE} \left[ \truepdf{Z} - \empiricalkdepdf{Z}\right]  \propto \left(10^{-1} \Nsamples^{-\frac{1}{5}}\right)^2.
\end{equation}
Thus, for instance, if we were to run our MC simulation sampling $\Nsamples=10^5$ samples, then we can expect the bias of $\empiricalkdepdf{Z}$ to be in the order of $10^{-4}$. This analysis on the bias allows us to consider the bias negligible if we are comparing it with quantities, such as $\pdfdist{}{\empiricalkdepdf{Z}(\Nsamples)}{\slpdf{Z}}$, order of magnitude greater than $10^{-4}$.
If this condition is met, then we can estimate the degree of approximation of binomial multiplication adopting the framework illustrated in Figure \ref{fig:ApproximationWithDistances} and instantiated for this specific SL operator as in Figure \ref{fig:ApproximationMultiplicationWithDistances}.

\subsection{Empirical Evaluation \label{sec:EmpircalStudy}}

\rev{In this section I added the analytic moment-matching and I changed the conclusions accordingly}{In this section} we describe our experimental simulations for the evaluation of the degree of approximation of the binomial product. We will first offer a qualitative analysis of the SL approximation $\slpdf{Z}$ and the approximation generated via MC simulation $\empiricalmcpdf{Z}$; then, we will provide a quantitative statistical assessment of the distance $\pdfdist{}{\truepdf{Z}}{\slpdf{Z}}$; next, we will analyze a specific study case concerning the worst-case scenario of the product of two degenerate Beta random variables; finally, we will assess quantitatively the degree of approximation in the product of multiple opinions.

In our simulations starting in the domain of subjective logic, opinions $\opinion{} = \binomialOp$ are sampled randomly. The parameters $b$, $d$ and $u$ must be sampled from a simplex defined by the constraint $b+d+u=1$; therefore we sample them from a Dirichlet distribution $\dirichletpdf{\pmb{\alpha}}$ with $\pmb{\alpha}=\left[1,1,1\right]$, which guarantees a uniform sampling over the simplex. The parameter $a$, instead, is sampled from a uniform distribution $\uniformpdf{0}{1}$. 
In the simulations starting in the domain of probability distributions, Beta distributions $\betapdf{\alpha}{\beta}$ are sampled randomly; both parameters $\alpha$ and $\beta$ are drawn from a uniform pdf on a bounded domain, $\uniformpdf{0}{10}$.

All the simulations are carried out using the WebPPL probabilistic programming language \cite{goodman2014design} and the scripts are available online\footnote{\url{https://github.com/FMZennaro/SLMC/BinomialProduct}}.

\subsubsection{Qualitative simulations \label{sec:Qualitative-Simulations}} 
In the qualitative simulations we aim at getting a first intuitive feeling about the approximation of $\slpdf{Z}$.

\paragraph{Protocol}
In order to compare the SL approximation $\slpdf{Z}$ and the MC approximation $\empiricalmcpdf{Z}$ we adopt the following protocol: (i) we sample two random opinions $\opinion{X}$ and $\opinion{Y}$; (ii) we compute their product $\opinion{Z} = \opinion{X} \multiplication \opinion{Y}$; (iii) we project $\opinion{Z}$ onto the distribution $\slpdf{Z}$; (iv) we project the opinions $\opinion{X}$ and $\opinion{Y}$ onto the Beta distributions $p_X$ and $p_Y$; (v) we re-create $\empiricalmcpdf{Z}$ using MC simulation to draw $\Nsamples$ samples $\lbrace z_1, z_2, \dots, z_\Nsamples \rbrace$ from $\truepdf{Z}$; finally, (vi) we plot $\slpdf{Z}$ against $\empiricalmcpdf{Z}$ numerically, without any smoothing or interpolation. 
On the side, (vii) we use the samples $\lbrace z_1, z_2, \dots, z_\Nsamples \rbrace$ to estimate moments via MC integration and then instantiate the moment-matching approximations $\empiricalgausspdf{Z}$ and $\empiricalbetapdf{Z}$; (viii) we use Equation \ref{eq:analyticMM} to compute the analytic moment-matching approximations $\mmgausspdf{Z}$ and $\mmbetapdf{Z}$; (ix) we plot the moment-matching approximations against $\empiricalmcpdf{Z}$ numerically.

\begin{figure}
\begin{centering}
	\includegraphics[scale=0.7]{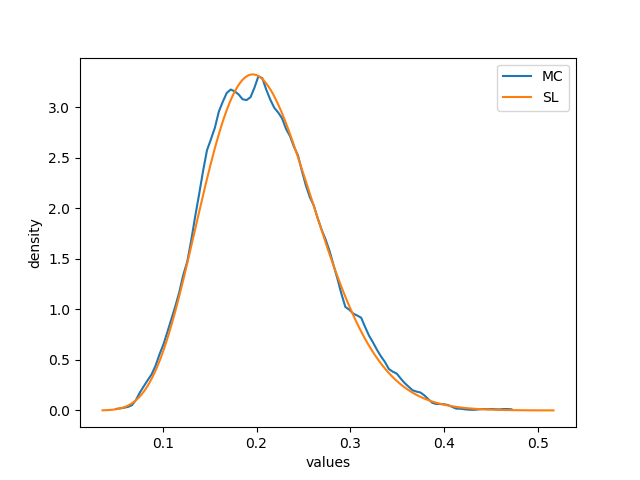}
\par\end{centering}
\caption{Qualitative simulation. The first opinion $\opinion{X}$ has parameters $b=0.61$, $d=0.30$, $u=0.09$ and $a=0.79$, the second opinion $\opinion{Y}$ has parameters $b=0.28$, $d=0.66$, $u=0.06$ and $a=0.46$. The number of samples is $\Nsamples=10^5$. \label{fig:qual1}}
\end{figure}
\begin{figure}
\begin{centering}
	\includegraphics[scale=0.6]{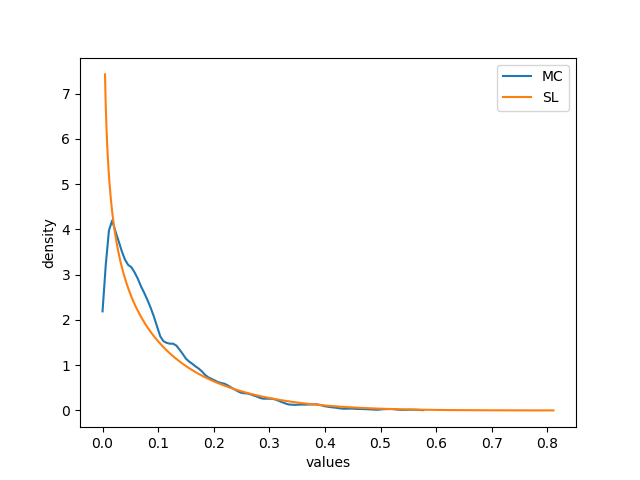}
\par\end{centering}

\caption{Qualitative simulation. The first opinion $\opinion{X}$ has parameters $b=0.16$, $d=0.55$, $u=0.29$ and $a=0.001$, the second opinion $\opinion{Y}$ has parameters $b=0.43$, $d=0.16$, $u=0.41$ and $a=0.39$. The number of samples is $\Nsamples=10^5$. \label{fig:qual2}}
\end{figure}

\paragraph{Results}
Figures \ref{fig:qual1} and \ref{fig:qual2} illustrates the difference between the SL binomial multiplication $\slpdf{Z}$ and the approximation of the true pdf $\truepdf{Z}$ plotted via MC. In some instances, $\slpdf{Z}$ seems to provide a very good approximation of $\truepdf{Z}$, as shown in Figure \ref{fig:qual1}. In other instances, as shown in Figure \ref{fig:qual2}, this approximation is more coarse, especially when it comes to values of the support near the extremes.

\begin{figure}
\begin{centering}
	\includegraphics[scale=0.6]{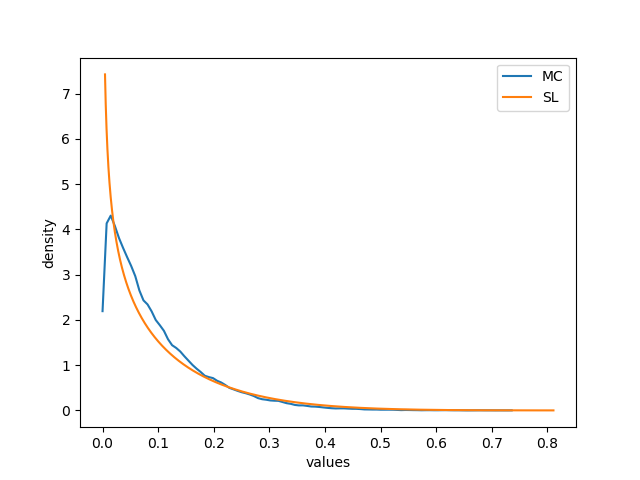}
\par\end{centering}
\caption{Qualitative simulation. Same settings as in Figure \ref{fig:qual2}, except for the number of samples $\Nsamples=10^6$. \label{fig:qual2_valid}}
\end{figure}

The discrepancy shown in Figure \ref{fig:qual2} may be theoretically imputed to a poor approximation of the MC simulation due to a limited number of samples. In order to confute this hypothesis, another identical simulation with a number of samples one order of magnitude larger was run. Figure \ref{fig:qual2_valid} shows that this simulation returned the same qualitative result. This suggests that the gap between $\empiricalmcpdf{Z}$ and $\slpdf{Z}$ may not be imputed to a poor MC approximation.

\begin{figure}
\begin{centering}
	\includegraphics[scale=0.8]{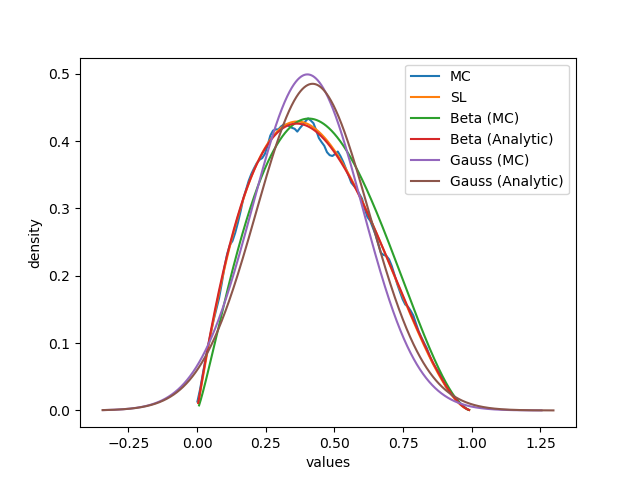}
\par\end{centering}
\caption{Qualitative simulation. The first opinion $\opinion{X}$ has parameters $b=0.35$, $d=0.23$, $u=0.42$ and $a=0.83$, the second opinion $\opinion{Y}$ has parameters $b=0.14$, $d=0.26$, $u=0.60$ and $a=0.77$. The number of samples is $\Nsamples=10^5$. \label{fig:qual3}}
\end{figure}

Figure \ref{fig:qual3} offers a visual comparison of the approximations offered by $\empiricalmcpdf{Z}$ and $\slpdf{Z}$ contrasted now with the Gaussian $\empiricalgausspdf{Z}$, $\mmgausspdf{Z}$ and the Beta $\empiricalbetapdf{Z}$, $\mmbetapdf{Z}$ approximations. The analytic and empirical (via MC) approximations behave in a very similar fashion. Overall, the Gaussian approximations are the farthest from $\truepdf{Z}$, while the Beta approximations follow very closely $\slpdf{Z}$ and $\empiricalmcpdf{Z}$; in particular, the analytic Beta approximation $\mmbetapdf{Z}$ almost overlap $\slpdf{Z}$, because of a similar approach in evaluating the moments of $\truepdf{Z}$.

\paragraph{Discussion}
This analysis suggests that the SL approximation $\slpdf{Z}$ may provide a good and useful estimation of the true pdf $\truepdf{Z}$; indeed, $\slpdf{Z}$ follows very closely the shape $\empiricalmcpdf{Z}$ which, in turn, is close to $\truepdf{Z}$. Given that $\slpdf{Z}$ consists of a smooth Beta distribution, it is not surprising that the approximation suffers the worst near the boundaries where the MC estimate $\empiricalmcpdf{Z}$ diverges from $\slpdf{Z}$, as shown in Figure \ref{fig:qual2}.
The results also discourage the naive possibility of using a Gaussian approximation, since $\empiricalgausspdf{Z}$ mismodels the true pdf $\truepdf{Z}$ by centering the mean but spreading probability mass too widely beyond the domain $[0,1]$. Instead, a Beta approximation $\empiricalbetapdf{Z}$ or $\mmbetapdf{Z}$ provides an approximation qualitatively very close to $\slpdf{Z}$ and $\empiricalmcpdf{Z}$; also, notice that the Beta approximation $\mmbetapdf{Z}$ can be computed as cheaply as the SL approximation, with complexity $\bigO{1}$, by evaluating its mean and variance analytically.

\subsubsection{Quantitative simulations \label{sec:Quantitative-Simulations}} 
Quantifying the gap between $\truepdf{Z}$ and $\slpdf{Z}$ that we observed in the qualitative study above is the aim of the quantitative simulations.

\paragraph{Protocol}
The first part of our quantitative protocol is the same as the qualitative protocol: 
(i-a) we sample two random opinions $\opinion{X}$ and $\opinion{Y}$; (ii-a) we compute their product $\opinion{Z} = \opinion{X} \multiplication \opinion{Y}$; (iii-a) we project $\opinion{Z}$ onto the distribution $\slpdf{Z}$; (iv-a) we project the opinions $\opinion{X}$ and $\opinion{Y}$ onto the Beta distributions $p_X$ and $p_Y$; (v-a) we re-create $\empiricalmcpdf{Z}$ using MC simulation to draw $\Nsamples$ samples $\lbrace z_1, z_2, \dots, z_\Nsamples \rbrace$ from $\truepdf{Z}$. 
Then, instead of plotting our results, (vi-a) we use a KDE to explicitly estimate $\empiricalkdepdf{Z}$; and, (vii-a) we compute via MC integration the area determined by the integral $\int_0^1 \absval{\slpdf{Z}(z)-\empiricalkdepdf{Z}(z)}dz$.
On the side, we compute moment-matching approximations as before: (viii) we use the samples $\lbrace z_1, z_2, \dots, z_\Nsamples \rbrace$ to estimate moments via MC integration and then instantiate  $\empiricalgausspdf{Z}$ and $\empiricalbetapdf{Z}$; (ix) we use Equation \ref{eq:analyticMM} to compute the analytic approximations $\mmgausspdf{Z}$ and $\mmbetapdf{Z}$; (x) we compute via MC integration the area determined by the absolute difference between $\empiricalkdepdf{Z}$ and each moment-matching approximation.

For completeness, we also run a simulation starting in the domain of pdfs: 
(i-b) we sample two random Beta pdfs $p_X$ and $p_Y$; (ii-b) we re-create $\empiricalmcpdf{Z}$ using MC simulation to draw $\Nsamples$ samples $\lbrace z_1, z_2, \dots, z_\Nsamples \rbrace$ from $\truepdf{Z}$; (iii-b) we use a KDE to explicitly estimate $\empiricalkdepdf{Z}$; (iv-b) we map the Beta distributions $p_X$ and $p_Y$ onto the opinions $\opinion{X}$ and $\opinion{Y}$; (v-b) we compute their product $\opinion{Z} = \opinion{X} \multiplication \opinion{Y}$; (vi-b) we project $\opinion{Z}$ onto the distribution $\slpdf{Z}$; and, (vii-b) we compute via MC integration the area determined by the integral $\int_0^1 \absval{\slpdf{Z}(z)-\empiricalkdepdf{Z}(z)}dz$.
As before, we also compute distances between $\empiricalkdepdf{Z}$ and empirical (via MC) and analytical moment-matching approximations as explained in the steps (viii)-(x) above.
 
In order to get significant statistical result, we repeat each simulation $100$ times and we compute the mean and the standard deviation of the distance $\empiricalpdfdist{I}{\truepdf{Z}}{\slpdf{Z}}$. 

Notice that, since the pdf $\empiricalkdepdf{Z}$ that we are trying to estimate is defined on a bounded interval, using a Gaussian kernel for KDE is a sub-optimal choice. The Gaussian kernel distributes the mass of probability over the entire real line, and thus we would inevitably spill part of the probability mass beyond the domain $[0,1]$. To solve this problem we adopt the \textit{logit trick} \cite{shalizi2013advanced}: instead of applying a Gaussian KDE to estimate $\empiricalkdepdf{Z}$ directly from the samples $\lbrace z_1, z_2, \dots, z_\Nsamples \rbrace$, we use a logit transform $\textnormal{logit}(x) = \log \frac{x}{1-x}$ to project the sample $\lbrace z_1, z_2, \dots, z_\Nsamples \rbrace$ onto the entire real line; we then apply a Gaussian KDE to the projected samples and rescale back the learned pdf to $\empiricalkdepdf{Z}$.


Refer to Figure \ref{fig:ApproximationMultiplicationWithDistances} for the diagram of the experimental protocol for the quantitative simulations.

\begin{figure}
	\begin{centering}
		\includegraphics[scale=0.8]{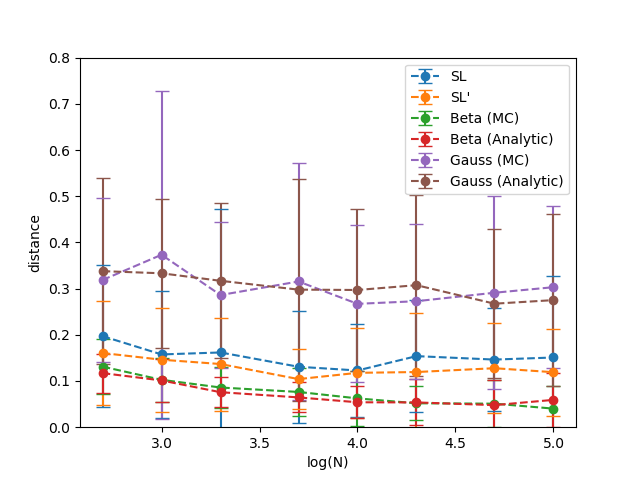}
		\par\end{centering}
	
	\caption{Quantitative simulation. Mean and standard deviation of the distance $\pdfdist{}{\empiricalkdepdf{Z}(\Nsamples)}{\cdot}$ when considering different approximations: $\slpdf{Z}$ in the simulation starting from opinions and following the steps (i-a)-(vii-a) (SL), $\slpdf{Z}$ in the simulation starting from Beta pdfs and following the steps (i-b)-(vii-b) (SL'), $\empiricalgausspdf{Z}$ (Gauss), $\empiricalbetapdf{Z}$ (Beta). Note that the $x$-axis is in a logarithmic (in base 10) scale. \label{fig:quant}}
\end{figure}

\paragraph{Results}
Figure \ref{fig:quant} shows the variation in the distances $\empiricalpdfdist{}{\truepdf{Z}}{\cdot}$ estimated as $\pdfdist{}{\empiricalkdepdf{Z}(\Nsamples)}{\cdot}$ as a function of the number samples $\Nsamples$ generated in the MC simulation. All the statistics are computed from $100$ repetitions and using $10^3$ uniformly sampled points on the support $[0,1]$ to perform MC integration.

The stable trend of all the distances $\pdfdist{}{\empiricalkdepdf{Z}(\Nsamples)}{\cdot}$ suggests that the MC simulations sampled enough points, for all the values of $\Nsamples$ that we considered, to return a good approximation. 

More importantly, recall that our whole analysis holds only if Equation \ref{eq:approximationTrueMC} is satisfied. Using $\Nsamples=10^5$, we know from Equation \ref{eq:kde_bias} that the bias in evaluating $\empiricalkdepdf{Z}(\Nsamples)$ is in the order of $10^{-4}$. Thus compared to the scale of the mean and variance error in our results, which are in the scale of $10^{-1}$, we can confirm that the bias is negligible. We can then state that $\pdfdist{}{\empiricalkdepdf{Z}(\Nsamples)}{\cdot}$ does indeed provide a good estimate of $\empiricalpdfdist{}{\truepdf{Z}}{\cdot}$.

Consistently with the previous experiments, the Gaussian approximations provides the worst approximation. Indeed, with distances $\pdfdist{}{\empiricalkdepdf{Z}(\Nsamples)}{\empiricalgausspdf{Z}}$ and $\pdfdist{}{\empiricalkdepdf{Z}(\Nsamples)}{\mmgausspdf{Z}}$  averaging around $0.30-0.35$, we can expect one sixth of the probability mass of a Gaussian approximation not to overlap with the true distribution $\truepdf{Z}$.

The Beta approximations $\empiricalbetapdf{Z}$ and $\mmbetapdf{Z}$ clearly offer a better solution. Even if the product of two Beta distributions $\truepdf{Z}$ is not a Beta distribution, it is clear from these results that the shape of $\truepdf{Z}$ is in general very close to a Beta pdf. Indeed, the expected value of $\pdfdist{}{\empiricalkdepdf{Z}(\Nsamples)}{\empiricalbetapdf{Z}}$ and $\pdfdist{}{\empiricalkdepdf{Z}(\Nsamples)}{\mmbetapdf{Z}}$ point out that $95\%$ of the mass of a Beta approximation and $\truepdf{Z}$ overlap with very limited variance.

The SL approximation $\slpdf{Z}$ also offers a good solution. The result of the simulation in which we started from opinions and the one in which we started from Beta pdfs are extremely close. This offers a confirmation of the robustness of the transformations between the domain of opinions and the domain of pdfs. Overall, the expected value of $\pdfdist{}{\empiricalkdepdf{Z}(\Nsamples)}{\slpdf{Z}}$ suggests that the typical overlap between the mass of $\slpdf{Z}$ and $\truepdf{Z}$ settles around $90\%$, slightly worse than the Beta approximation. The high variance points to a strong case-by-case variability: in certain scenario $\slpdf{Z}$ may provide a model as good or better than $\empiricalbetapdf{Z}$ or $\mmbetapdf{Z}$, but on other instances its quality may degrade further.

%

\paragraph{Discussion}
The results of our quantitative analysis agree with the qualitative study. A Gaussian approximation $\empiricalgausspdf{Z}$ or $\mmgausspdf{Z}$ was shown to be a poor choice for modelling the product of two Beta distributions (and, for this reason, we will drop this approximation from the next simulations). Instead, the SL approximation $\slpdf{Z}$ and the Beta approximations $\empiricalbetapdf{Z}$ or $\mmbetapdf{Z}$ are both good approximations, assuming that we can accept a difference between the true pdf and the approximation up to $5\%-10\%$ of the probability density. With limited computational resources, $\mmbetapdf{Z}$ seems to be, on average, the best bet.

\subsubsection{Limit-case Study}
In this limit-case study, we consider the worst-case scenario considered in \cite{josang2005multiplication}. This study provides a way to enrich the previous study and reconnect this paper to it.

\paragraph{Protocol}
We quantitatively analyze the case in which both opinions $\opinion{X}$ and $\opinion{Y}$ are degenerate Beta pdfs of the form $\betapdf{1}{1}$ with $a=\frac{1}{2}$. 
To provide a quantitative analysis we follow the same protocol used in Section \ref{sec:Quantitative-Simulations}: first, we derive the MC approximation $\empiricalkdepdf{Z}$ (using the KDE algorithm), the SL approximation $\slpdf{Z}$, the empirical (via MC) Beta approximation $\empiricalbetapdf{Z}$ and the analytic Beta approximation $\mmbetapdf{Z}$; then, we compute the distance between the aforementioned distributions and the true pdf $\truepdf{Z}$, whose exact form, $-\log(z)$, is given in \cite{josang2005multiplication}.

\paragraph{Results} Table \ref{tab:case1} shows the evaluation of the distance $\empiricalpdfdist{}{\truepdf{Z}}{\cdot}$ with respect to the true pdf $\truepdf{Z} = -\log(z)$, when performing MC simulations with $\Nsamples=10^6$ points and using $10^3$ uniformly sampled points on the support $[0,1]$ to perform MC integration.

The results show that the difference between the approximations is about one order of magnitude from each other. The MC approximation is, as expected, very close to the true pdf $\truepdf{Z}$, with a distance averaging around $9\cdot10^{-3}$. This is higher than the theoretically computed value, likely due to the fact that we are evaluating a limit case; however, this difference is still small enough to allow us a comparison with the other approximations. The Beta approximations have a slightly higher distance around $3\cdot10^{-2}$. Finally, the SL approximation has the highest distance at around $2\cdot10^{-1}$, meaning that the probability mass of $\slpdf{Z}$ and $\truepdf{Z}$ overlap for about $90\%$. All the results also show a high variance, which is caused by the difficulty in numerically approximating values near $0$, where the true pdf $\truepdf{Z} = -\log(z) $ diverges.

\begin{table}
	\begin{centering}
	\begin{tabular}{cc}
		\hline 
		& $\empiricalpdfdist{}{\truepdf{Z}}{\cdot}$\tabularnewline
		\hline 
		KDE & 0.00871 $\pm$ 0.01242\tabularnewline
		\hline 
		SL & 0.20793 $\pm$ 0.93301\tabularnewline
		\hline 
		Beta (MC) & 0.03070 $\pm$ 0.10018\tabularnewline
		\hline
		Beta (Analytic) & 0.03635 $\pm$ 0.14617\tabularnewline
		\hline  
	\end{tabular}
	\par\end{centering}	
	\caption{Limit case study. Evaluation of the distance $\empiricalpdfdist{}{\truepdf{Z}}{\cdot}$ when using a MC approximation $\empiricalmcpdf{Z}$, a SL approximation $\slpdf{Z}$ and an empirical Beta approximation $\empiricalbetapdf{Z}$, and an analytic Beta approximation $\mmbetapdf{Z}$. \label{tab:case1}}
\end{table}

\paragraph{Discussion} The results are consistent with our previous results obtained in the quantitative analysis in Section \ref{sec:Quantitative-Simulations} and they confirm that the scenario considered in \cite{josang2005multiplication} with two degenerate Beta pdfs of the form $\betapdf{1}{1}$ and $a=\frac{1}{2}$ is indeed a hard case for SL approximation. The MC approximation performs better in modeling the true form of the pdf $-\log(z)$ and, compared to it, the SL approximation is two orders of magnitude less precise in terms of integral distance. This simulation thus clearly highlights the cost in terms of accuracy that the computational simplicity of SL implies.

\subsubsection{Multiple Products}
In this last experimental section we consider the product of multiple opinions and we examine how approximation spread.

\paragraph{Protocol}
We quantitatively evaluate the product of multiple opinions $\opinion{X_1}$, $\opinion{X_2}$ ... $\opinion{X_L}$ by randomly sampling $L$ opinions and then defining the product $\opinion{Z} = \left( \left( \left( \opinion{X_1} \multiplication \opinion{X_2} \right) \multiplication \opinion{X_3} \right) \dots \multiplication \opinion{X_L}\right)$ and $\truepdf{Z} = p_{X_1} \cdot p_{X_2} \dots \cdot p_{X_L} $. The following analysis adopts the same protocol used in the quantitative simulations in Section \ref{sec:Quantitative-Simulations} in order to compute the SL approximation $\slpdf{Z}$ and the analytic Beta approximation $\mmbetapdf{Z}$, and then evaluate the integral distances $\empiricalpdfdist{I}{\truepdf{Z}}{\cdot} \estimate \pdfdist{}{\empiricalkdepdf{Z}(\Nsamples)}{\cdot}$, where now the final pdf over $Z$ is given by the product of multiple opinions. Notice that, while the convergence properties of the MC simulation remains the same, we may expect the precision of the SL approximation and the Beta approximation to degrade over multiple products as successive approximations cumulate.

\begin{figure}
	\begin{centering}
		\includegraphics[scale=0.8]{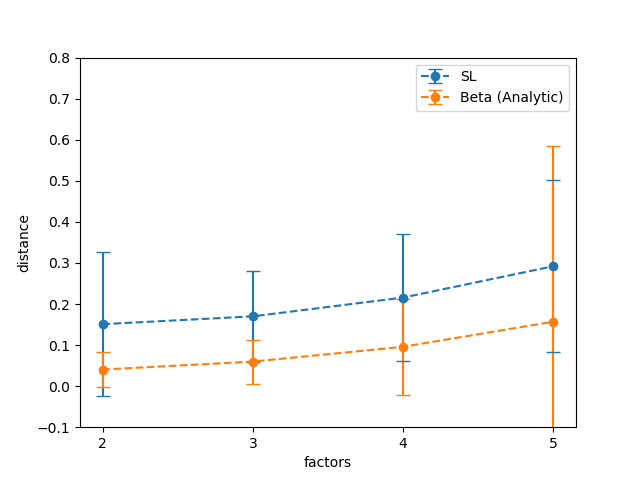}
		\par\end{centering}
	
	\caption{Multiple products. Mean and standard deviation of the distance $\pdfdist{}{\empiricalkdepdf{Z}(\Nsamples)}{\slpdf{Z}}$ and $\pdfdist{}{\empiricalkdepdf{Z}(\Nsamples)}{\mmbetapdf{Z}}$  when the distribution $Z$ is given by the product of multiple factors. \label{fig:multiple}}
\end{figure}

\paragraph{Results} Figure \ref{fig:multiple} shows the variation of the distance $\pdfdist{}{\empiricalkdepdf{Z}(\Nsamples)}{\slpdf{Z}}$ and $\pdfdist{}{\empiricalkdepdf{Z}(\Nsamples)}{\mmbetapdf{Z}}$ as a function of the number $L$ of opinions that are multiplied together to determine $Z$. A slight increase in the degree of approximation may be observed as the number of factors increases from $2$ to $5$.

\paragraph{Discussion}
The hypothesis that the degree of approximation of the SL operator and analytical Beta approximation degrades over multiple products because of the accumulation of approximation appears to be correct. As more factors are taken into consideration, both $\slpdf{Z}$ and $\mmbetapdf{Z}$ slowly diverges from the true pdf $\truepdf{Z}$. A modeler should be aware of these dynamics in case she were to use SL to approximate the product of multiple Beta random variables.

\section{Case Study: Fusion of Beta Distributions \label{sec:CFusion}}

\rev{This whole section is new}{In this section}, we further showcase the versatility of our framework by applying it to yet another subjective logic operator. We first provide the definition of fusion of Beta distributions; we then introduce the SL operator for fusion and we discuss its approximation and complexity; finally, we apply our framework to the problem of getting an approximation of fusion and we run empirical simulations to validate our theoretical results. 

\paragraph{Fusion of Beta pdfs} Let $X \distributes \betapdf{\alpha_X}{\beta_X}$ and $Y \distributes \betapdf{\alpha_Y}{\beta_Y}$ be two independent Beta random variables with associated pdfs $p_X$ and $p_Y$. Let us define a third random variable $Z$ as the fusion of the two Beta random variables $Z = X \cfusion Y$ \cite{kaplan2018efficient}:
\[
X \cfusion Y = \frac{x \multiplication y}{x \multiplication y + (1-x) \multiplication (1-y)},
\]
where $x$ and $y$ are realizations of the random variables $X$ and $Y$.
As in the case of the product of Beta random variables, determining the shape of $p_Z$ is a non-trivial problem. MC simulations offer a robust method to sample points from the probability distribution of $Z$ and to estimate $\truepdf{Z}$ by moment-matching or kernel density estimation. Notice that, differently from the previous case study, we do not have an exact analytical solution for computing the first two moments of $\truepdf{Z}$ \cite{kaplan2018efficient}.

\paragraph{Subjective logic fusion}
A SL operator may be instantiated to compute an approximate fusion over two binomial opinions.
Given two binomial opinions $\opinion{X}$ and $\opinion{Y}$ defined on the same domain $\Omega$ and with the same prior $a$, we define the binomial opinion $\opinion{Z}$ resulting from the fusion $\opinion{X} \cfusion \opinion{Y}$ as:
\begin{equation} \label{eq:SL_CFusion}
\opinion{Z} = 
\begin{cases}
b_{Z}=\frac{m_{Z}s_{Z}-Wa}{s_{Z}}\\

d_Z = \frac{(1-m_{Z})s_{Z}-W(1-a)}{s_{Z}}\\

u_Z  = \frac{W}{s_{Z}}\\

a_Z = a,
\end{cases}
\end{equation}
where
\footnotesize
\begin{equation} \label{eq:SL_CFusion1}
m_{Z}=\frac{b_{X}b_{Y}+b_{Y}au_{X}+b_{X}au_{Y}+a^{2}u_{X}u_{Y}}{2\left(b_{X}b_{Y}+b_{Y}au_{X}+b_{X}au_{Y}+a^{2}u_{X}u_{Y}\right)+1-b_{Y}-au_{Y}-b_{X}-au_{X}}\\
\end{equation}
\begin{equation*} 
\begin{split}
s_{Z}=&\max\left\{\frac{Wa}{m_{Z}},\frac{W(1-a)}{(1-m_{Z})}, \right. \\
& \left. \frac{\left(\frac{W}{u_{X}}+1\right)\left(\frac{W}{u_{Y}}+1\right)\left(b_{X}-b_{X}^{2}-au_{X}-a^{2}u_{X}^{2}\right)\left(b_{Y}-b_{Y}^{2}-au_{Y}-a^{2}u_{Y}^{2}\right)}{m_{Z}(1-m_{Z})\left[\left(\frac{W}{u_{Y}}+1\right)\left(b_{Y}-b_{Y}^{2}-au_{Y}-a^{2}u_{Y}^{2}\right)+\left(\frac{W}{u_{X}}+1\right)\left(b_{X}-b_{X}^{2}-au_{X}-a^{2}u_{X}^{2}\right)\right]}-1\right\}.
\end{split}
\end{equation*}
\normalsize
This formula expresses in the subjective logic formalism the moment-matching approximation of fusion defined in \cite{kaplan2018efficient}.
Practically, a fusion operator allows us to evaluate the aggregation of two different opinions on the same fact.

\paragraph{Approximating the product of Beta pdfs}
Now, if we are interested in computing the fusion $Z = X \cfusion Y$, where $X$ and $Y$ are two independent Beta random variables, we may rely either on the SL approximation defined in Equation \ref{eq:SL_CFusion} or on approximations computed via MC simulations; as before we will consider the following approximations for the true pdf $\truepdf{Z}$: a SL pdf $\slpdf{Z}$, a KDE estimation $\empiricalkdepdf{Z}$, a Beta and a Gaussian moment-matching approximation $\empiricalbetapdf{Z}$, $\empiricalgausspdf{Z}$ computed by evaluating mean and variance of $\truepdf{Z}$ via MC integration. Differently from before, we do not consider an exact analytical moment-matching approximation ($\mmbetapdf{Z}$ or $\mmgausspdf{Z}$) because no exact analytical formula exists. Notice, also, that since we used the moment-matching approximation provided in \cite{kaplan2018efficient} to define the SL operator in Equation \ref{eq:SL_CFusion}, our results on the degree of approximation of the SL operator immediately extend to Operator 1 defined in \cite{kaplan2018efficient}. 

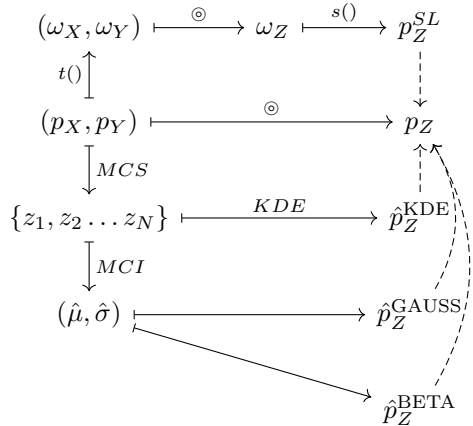
\begin{figure}
	\centering
	\begin{tikzcd}
		\left(\opinion{X}, \opinion{Y} \right) \arrow[mapsto]{r}{\cfusion} 
		& \opinion{Z} \arrow[mapsto]{r}{s()} & \slpdf{Z}  \arrow[dashrightarrow]{d} \\		
		\left(p_X, p_Y\right) \arrow[mapsto]{rr}{\cfusion} \arrow[mapsto]{u}{t()} \arrow[mapsto]{d}{MCS} & & \truepdf{Z}  \\
		\lbrace z_1, z_2 \dots z_\Nsamples \rbrace  \arrow[mapsto]{d}{MCI} \arrow[mapsto]{rr}{KDE} &  &
		\empiricalkdepdf{Z} \arrow[dashrightarrow]{u} \\
		\left( \hat{\mu}, \hat{\sigma} \right) \arrow[mapsto]{rr} \arrow[mapsto]{drr} & & \empiricalgausspdf{Z} \arrow[dashrightarrow, bend right]{uu} \\
		& & \empiricalbetapdf{Z} \arrow[dashrightarrow, bend right]{uuu}
	\end{tikzcd}
	\caption{Approximations of the fusion of two Beta pdfs, $p_X$ and $p_Y$. \textit{MCS} stands for MC sampling, \textit{MCI} stands for MC integration, \textit{KDE} stands for kernel-density estimation. \label{fig:ApproximationCFusion}}
\end{figure}

Figure \ref{fig:ApproximationCFusion} provides the concrete instantiation of the diagram in Figure \ref{fig:Allapproximation} for the SL operator of fusion and illustrates the alternative between the path of MC simulations and SL approximation.

\paragraph{Computational complexity of the fusion operator}
Despite the more involved expressions in Equation \ref{eq:SL_CFusion} and \ref{eq:SL_CFusion1}, the computational complexity for evaluating $\opinion{Z}$ is still constant with respect to the given opinions $\opinion{X}$ and $\opinion{Y}$; as such, the asymptotic complexity of the SL operator for fusion is $\bigO{1}$.

\paragraph{Approximation of the binomial operator}
The degree of approximation of the SL operator for fusion is equivalent to the approximation of the moment-matching solution on which it is defined; \cite{kaplan2018efficient} offers an evaluation of this approximation within the wider context of belief propagation in second-order Bayesian networks. In the following paragraphs, we will instead aim at estimating, in a more directed way, the approximation of the SL operator via the computation of the distance $\empiricalpdfdist{I}{\truepdf{Z}}{\slpdf{Z}}$.

\paragraph{Relating the binomial multiplication and Monte Carlo approximations}
Following the approach described in the previous section, we will compute a numerical estimation of the degree of approximation of the SL operator for fusion relying on the framework described in Section \ref{sec:Framework}.

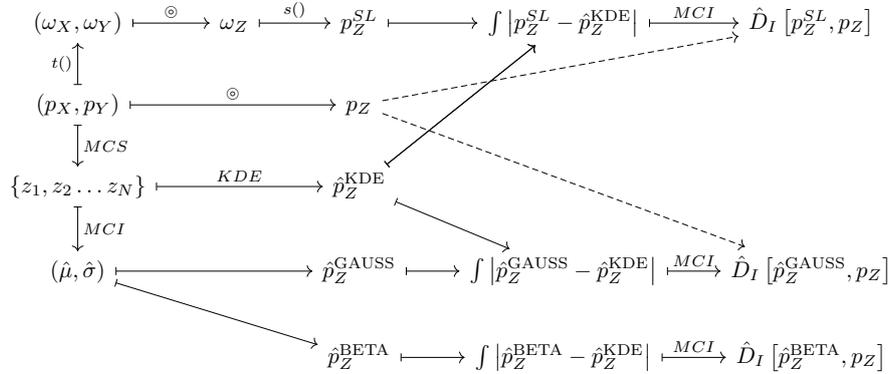
\begin{figure}
	\centering
	\adjustbox{scale=.85}{
	\begin{tikzcd}
		\left(\opinion{X}, \opinion{Y} \right) \arrow[mapsto]{r}{\cfusion} 
		& \opinion{Z} \arrow[mapsto]{r}{s()} & \slpdf{Z} \arrow[mapsto]{r} & \int \absval{\slpdf{Z} - \empiricalkdepdf{Z}} \arrow[mapsto]{r}{MCI} & \empiricalpdfdist{I}{\slpdf{Z}}{\truepdf{Z}}\\		
		\left(p_X, p_Y\right) \arrow[mapsto]{rr}{\cfusion} \arrow[mapsto]{u}{t()} \arrow[mapsto]{d}{MCS} & & \truepdf{Z} \arrow[dashrightarrow]{urr} \arrow[dashrightarrow]{ddrr} \\
		\lbrace z_1, z_2 \dots z_\Nsamples \rbrace  \arrow[mapsto]{d}{MCI} \arrow[mapsto]{rr}{KDE} &  &
		\empiricalkdepdf{Z} \arrow[mapsto]{uur} \arrow[mapsto]{dr} \arrow[mapsto]{uur}\\
		\left( \hat{\mu}, \hat{\sigma} \right) \arrow[mapsto]{rr} \arrow[mapsto]{drr} & & \empiricalgausspdf{Z} \arrow[mapsto]{r} & \int \absval{\empiricalgausspdf{Z} - \empiricalkdepdf{Z}} \arrow[mapsto]{r}{MCI} & \empiricalpdfdist{I}{\empiricalgausspdf{Z}}{\truepdf{Z}} \\
		& & \empiricalbetapdf{Z} \arrow[mapsto]{r} & \int \absval{\empiricalbetapdf{Z} - \empiricalkdepdf{Z}} \arrow[mapsto]{r}{MCI} & \empiricalpdfdist{I}{\empiricalbetapdf{Z}}{\truepdf{Z}}
	\end{tikzcd}}
	\caption{Evaluations of the distances between $\truepdf{Z}$ and its approximations based on the assumption that $\pdfdist{}{\truepdf{Z}}{\empiricalkdepdf{Z}(\Nsamples)} \approx 0$. \textit{MCS} stands for MC sampling, \textit{MCI} stands for MC integration, \textit{KDE} stands for kernel-density estimation. \label{fig:ApproximationCFusionWithDistances}}
\end{figure}

Once again, we need to check that the basic condition expressed in Equation \ref{eq:approximationTrueMC} requiring the bias of $\empiricalkdepdf{Z}$ to be negligible is satisfied. Given that we will compute KDE using a Gaussian kernel with width defined by the Silverman rule (Equation \ref{eq:Silverman}), and given that the support of $Z$ is bounded on $\left[0,1\right]$, we can again expect the bias of our KDE estimator to be in the order of $\left(10^{-1} \Nsamples^{-\frac{1}{5}}\right)^2$ (Equation \ref{eq:kde_bias}). Thus, the KDE bias may be considered negligible in the estimation of distances, if the distances we are considering are orders of magnitude greater than this bias. If this condition is met, then we can estimate the degree of approximation of fusion adopting the framework defined in Figure \ref{fig:ApproximationWithDistances} and instantiated in Figure \ref{fig:ApproximationCFusionWithDistances}.

\subsection{Empirical Evaluation}

In this section we describe our experimental simulations for the evaluation of the degree of approximation of fusion. We will first provide a qualitative assessment of the SL approximation $\slpdf{Z}$ and the approximations generated via MC simulations; then, we will provide a quantitative statistical evaluation of the distance $\pdfdist{}{\truepdf{Z}}{\cdot}$ for the SL approximation and for Beta and Gaussian moment-matching approximations.

In our simulations we randomly generate opinion and pdfs using the same protocol defined in Section \ref{sec:EmpircalStudy}. These simulations are also run using the WebPPL probabilistic programming language \cite{goodman2014design} and the scripts are available online\footnote{\url{https://github.com/FMZennaro/SLMC/Fusion}}.

\subsubsection{Qualitative simulations} 
In the qualitative simulations we try to offer a first visual assessment of the quality of the approximation of $\slpdf{Z}$.

\paragraph{Protocol}
We follow the same protocol defined for the qualitative simulations in Section \ref{sec:EmpircalStudy}, just replacing the operations for binomial multiplication with the operations for fusion.

\begin{figure}
	\begin{centering}
		\includegraphics[scale=0.8]{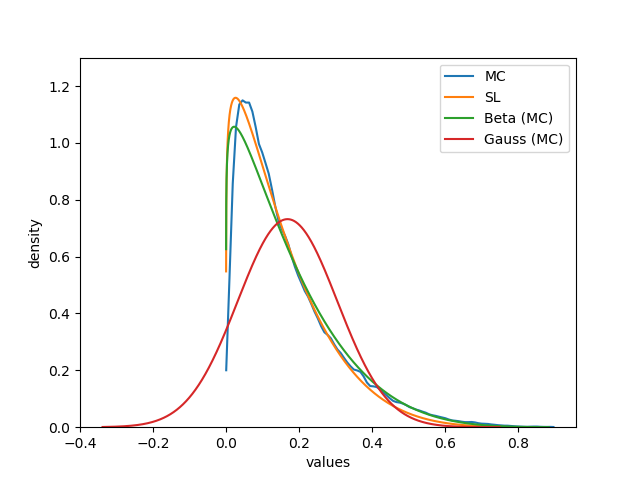}
		\par\end{centering}
	
	\caption{Qualitative simulation. The first opinion $\opinion{X}$ has parameters $b=0.16$, $d=0.58$, $u=0.26$ and $a=0.57$, the second opinion $\opinion{Y}$ has parameters $b=0.18$, $d=0.64$, $u=0.18$ and $a=0.57$. The number of samples is $\Nsamples=10^6$. \label{fig:qual1_CF}}
\end{figure}

\begin{figure}
	\begin{centering}
		\includegraphics[scale=0.8]{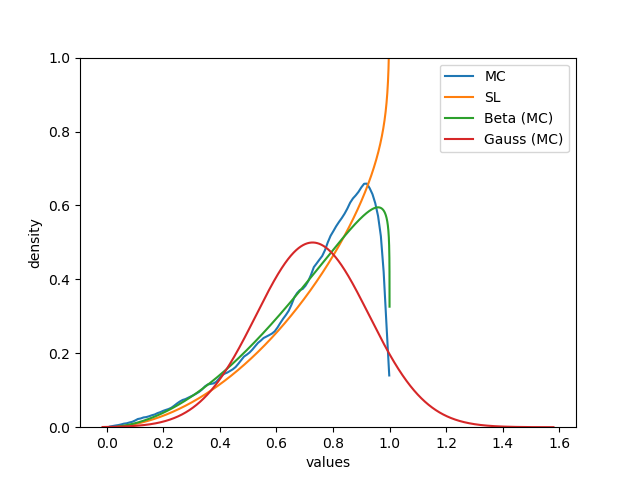}
		\par\end{centering}
	
	\caption{Qualitative simulation. The first opinion $\opinion{X}$ has parameters $b=0.14$, $d=0.63$, $u=0.23$ and $a=0.54$, the second opinion $\opinion{Y}$ has parameters $b=0.83$, $d=0.05$, $u=0.12$ and $a=0.54$. The number of samples is $\Nsamples=10^6$. \label{fig:qual2_CF}}
\end{figure}

\paragraph{Results}
Figures \ref{fig:qual1_CF} and \ref{fig:qual2_CF} offer a visual comparison of the approximations offered by $\empiricalmcpdf{Z}$ and $\slpdf{Z}$ along with the Gaussian $\empiricalgausspdf{Z}$ and the Beta $\empiricalbetapdf{Z}$ moment-matching approximations computed via MC integration. 
While the pdf underlying the fusion of two random variables is not Beta distributed, both the SL approximation $\slpdf{Z}$ and the moment-matching approximation $\empiricalbetapdf{Z}$ seem to offer a good coarse approximation of the true distribution $\truepdf{Z}$; both approximations match very well the true distribution within the support $[0,1]$, but they may show problems modelling the behaviour of $\truepdf{Z}$ near the boundaries of the domain. As before, the Gaussian approximation $\empiricalgausspdf{Z}$ offers the worst match, as its shape is not ideal to model a pdf on a bounded domain.

\paragraph{Discussion}
This qualitative assessment suggests that the SL approximation $\slpdf{Z}$ may provide a good enough approximation of the true pdf $\truepdf{Z}$ at a very low computational cost. The Beta moment-matching approximation evaluated computing mean and variance via MC simulations seems to behave in a very similar fashion; however, in this case where no exact analytical moment-matching approximation is possible, the cost of evaluating $\empiricalbetapdf{Z}$ corresponds to the cost of running the whole MC simulation. Last, the Gaussian moment-matching approximation $\empiricalgausspdf{Z}$ constitutes a sub-optimal choice, as its fit is worse than the alternative approximations and its computational cost is equivalent to the Beta moment-matching approximation.

\subsubsection{Quantitative simulations} 
We now move to quantifying the gap between $\truepdf{Z}$ and $\slpdf{Z}$ observed in the above simulations.

\paragraph{Protocol}
We follow the same protocol defined for the quantitative simulations in Section \ref{sec:EmpircalStudy}, just replacing the operations for binomial multiplication with the operations for fusion. As before, Figure \ref{fig:ApproximationMultiplicationWithDistances} offers a diagram of our experimental protocol.

\begin{figure}
	\begin{centering}
		\includegraphics[scale=0.8]{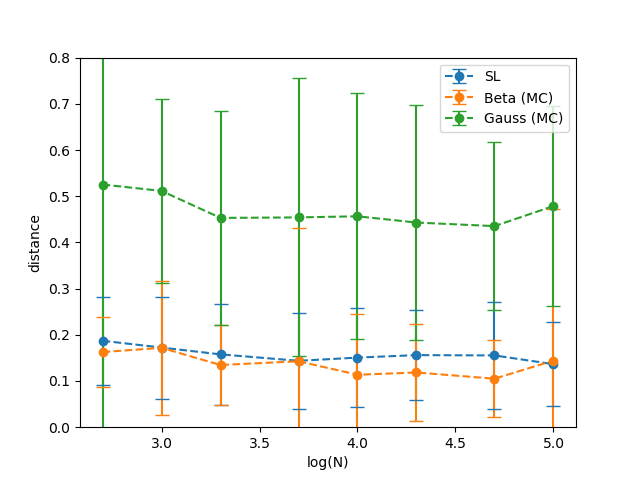}
		\par\end{centering}
	
	\caption{Quantitative simulation. Mean and standard deviation of the distance $\pdfdist{}{\empiricalkdepdf{Z}(\Nsamples)}{\cdot}$ when considering different approximations: $\slpdf{Z}$ (SL), $\empiricalgausspdf{Z}$ (Gauss) and $\empiricalbetapdf{Z}$ (Beta). Note that the $x$-axis is in a logarithmic (in base 10) scale. \label{fig:quantCF}}
\end{figure}

\paragraph{Results}
Figure \ref{fig:quantCF} presents the estimation of the distances $\empiricalpdfdist{}{\truepdf{Z}}{\cdot}$ as a function of the number samples $\Nsamples$ generated in the MC simulation. These statistics are computed performing $100$ repetitions and using $10^3$ uniformly sampled points on the support $[0,1]$ for MC integration.

In general, as in the previous simulation, we can make two preliminary observation: (i) all the distances show a stable trend, thus suggesting that our MC results are approaching their asymptotic limit; (ii) the order of magnitude of the distances is significantly greater than the KDE bias (Equation \ref{eq:kde_bias}), thus meaning that its bias is negligible with respect to the distances.

Confirming the previous qualitative investigation, the distance between the true pdf $\truepdf{Z}$ and the Gaussian approximation $\empiricalgausspdf{Z}$ reaches values as high as $0.5$, meaning that up to one fourth of the probability mass of $\empiricalgausspdf{Z}$ does not to overlap with the true distribution $\truepdf{Z}$.

The Beta approximation $\empiricalbetapdf{Z}$ models the true pdf $\truepdf{Z}$ better, with a distance around $0.10-0.15$, suggesting a good approximation in which $95\%$ of the mass of $\empiricalbetapdf{Z}$ overlaps with $\truepdf{Z}$.

Similarly, the SL approximation $\slpdf{Z}$ provides an equally good solution. The values of distance for $\slpdf{Z}$ are well within the range of the standard deviation of $\empiricalbetapdf{Z}$, suggesting that the degree of approximation of SL and Beta moment-matching are very close.

\paragraph{Discussion}
The results of this analysis agree with the previous qualitative simulations. Moreover, even if these results are quantitatively different, they are qualitatively in line with our study of the binomial product operator. The quantitative difference between the approximation of the binomial product and the fusion may be likely ascribed to the fact that it is easier to model the product of independent random variables instead of an arbitrary operation like fusion. From a qualitative point of view, though, the Gaussian approximation $\empiricalgausspdf{Z}$ ranks again last among the modeling options, while the SL approximation $\slpdf{Z}$ and the Beta approximations $\empiricalbetapdf{Z}$ offer better solutions, at a computational cost that is constant (in the case of SL) or linear in the number of MC samples (in the case of Beta approximation).

\section{Conclusion and Future Work \label{sec:Conclusion}}

In this paper we studied the use of subjective logic as a framework for approximating operations over probability distributions. As in the case of any approximation, we considered SL operators from the perspective of the trade-off between the computational simplicity they guarantee and the precision they sacrifice. We proposed a protocol based on MC simulations to evaluate quantitatively this trade-off, estimating the distance between the SL approximation and a KDE estimation, under the assumption of a negligible bias between the KDE reconstruction and the true probability distribution. 

\rev{Edited this part of the conclusion}{We applied our protocol} to the case study of the product and the fusion of two independent Beta distributions. The first case is relevant to fields like reliability analysis, while the second one is used in the field of subjective logic.
In general, SL operators guarantee the preservation of the first moment, but do not strictly preserve higher moments or quantiles. To quantify the degree of approximation of the SL operators, we compared them with other standard approximations, such as moment-matching with a Gaussian pdf, moment-matching with a Beta pdf, and KDE via MC.
Our simulations showed that, at the cost of accepting a difference between the SL approximation and the true pdf, SL offers a computationally efficient approximation. Both in the case of binomial products and in the case of fusion, the degree of approximation can be quantified in a mismatch between the SL approximation and the true pdf of up to $10\%$ of the probability mass.
In general, KDE approximation and Beta approximation provided better estimation; KDE, however, has a computational cost that is quadratic in the number of samples generated via MC; moment-matching has a computational cost that can be constant and comparable to SL when moments of interest can be computed analytically, or, otherwise, linear in the number of samples generated via MC. 

In summary, it is possible to enjoy the computational efficiency and the interpretability of SL if the modeling scenario allows room for approximations up to the amount estimated using our protocol. The recommendation is that, were SL operators to be used to model critical systems (as in the case of reliability analysis or when higher-order moments are critical), this divergence between the true pdf and the SL approximation that we highlighted should be factored in the analysis.

Further work will be developed for better characterizing the difference between true pdfs and SL approximations; in particular, understanding how the mass is differently allocated with respect to the overall shape of the pdf, whether, for instance, these differences are more accentuated near the mode (assuming one exists) or around the tail. According to the way in which probability mass is misplaced in SL approximations different forms of correction may be then considered. 

\paragraph{Acknowledgment}
The authors would like to express their gratitude to the anonymous reviewers who commented on the first version of this article and helped improving it.

\bibliographystyle{plain}
\bibliography{../../lib/lib}

\end{document}